\documentclass[aps,pra,reprint,showpacs,nofootinbib,superscriptaddress]{revtex4-1}
\input{Qcircuit}
\usepackage{graphicx,comment}
\usepackage{amsfonts}
\usepackage{amsmath,amssymb}
\usepackage{nth}
\usepackage{bm}
\usepackage{color}
\usepackage{epstopdf}
\usepackage{epsfig}
\usepackage{tikz}
\usepackage{pgfplots}
\pgfplotsset{
table/search path={csv/},
}
\usetikzlibrary{pgfplots.groupplots}
\usepackage[font=small,
   justification=justified,
   format=plain]{caption} % 'format=plain' avoids hanging indentation
\usepackage{subfig}
\usepackage{verbatim}
\usepackage{hyperref}
\usepackage{multirow}
\usepackage{tabularx}
\usepackage{wrapfig}
\hypersetup{
     colorlinks   = true,
    citecolor    = blue
}
\usepackage{lineno}
\usepackage[thicklines]{cancel}
\usepackage{color, colortbl}
\definecolor{LightCyan}{rgb}{0.88,1,1}
\usepackage{url}
    
{\rm }

\def\be{\begin{equation}}
 \def\ee{\end{equation}}
 \def\bea{\begin{eqnarray}}
 \def\eea{\end{eqnarray}}
\def\2{\frac{1}{2}}
\def\4{\frac{1}{4}}

\begin{document}
% \pagestyle{fancy}
% \rhead{\includegraphics[width=2.5cm]{vch-logo.png}}

\footnotetext{This manuscript has been authored by UT-Battelle, LLC, under Contract No. DE-AC0500OR22725 with the U.S. Department of Energy. The United States Government retains and the publisher, by accepting the article for publication, acknowledges that the United States Government retains a non-exclusive, paid-up, irrevocable, world-wide license to publish or reproduce the published form of this manuscript, or allow others to do so, for the United States Government purposes. The Department of Energy will provide public access to these results of federally sponsored research in accordance with the DOE Public Access Plan.}%

%\title{Studying Chemistry Benchmark Using VQE vs. QITE and Quantum Lanczos}
\title{Benchmarking Quantum Chemistry Computations with Variational, Imaginary Time Evolution, and Krylov Space Solver Algorithms}  
  
\author{K\"ubra Yeter-Aydeniz }
\email{yeteraydenik@ornl.gov}
\affiliation{Physics Division, Oak Ridge National Laboratory,
  Oak Ridge, TN 37831, USA}
\affiliation{Computational Sciences and Engineering Division, Oak Ridge National Laboratory,
  Oak Ridge, TN 37831, USA}

\author{Bryan T. Gard}
\email{bgard1@vt.edu}
\affiliation{Department of Physics, Virginia Tech, Blacksburg, VA 24061, U.S.A}

\author{Jacek Jakowski }
\email{jakowskij@ornl.gov}
\affiliation{Computational Sciences and Engineering Division, Oak Ridge National Laboratory,  Oak Ridge, TN 37831, USA}
  
\author{Swarnadeep Majumder}\email{swarnadeep.majumder@duke.edu}
\affiliation{Department of Electrical and Computer Engineering, Duke University, Durham, NC 27708 USA}%

\author{George S. Barron}
\email{gbarron@vt.edu}
\affiliation{Department of Physics, Virginia Tech, Blacksburg, VA 24061, U.S.A}

\author{George Siopsis}
\email{siopsis@tennessee.edu}
\affiliation{Department of Physics and Astronomy,  The University of Tennessee, Knoxville, TN 37996-1200, USA}

\author{Travis Humble}
\email{humblets@ornl.gov}
\affiliation{Quantum Science Center, Oak Ridge National Laboratory, Oak Ridge, Tennessee 37831-6211 USA}

\author{Raphael C.\ Pooser}
\email{pooserrc@ornl.gov}
\affiliation{Computational Sciences and Engineering Division, Oak Ridge National Laboratory,
  Oak Ridge, TN 37831, USA}

\date{\today}

\begin{abstract}
The rapid progress of noisy intermediate-scale quantum (NISQ) computing underscores the need to test and evaluate new devices and applications. Quantum chemistry is a key application area for these devices, and therefore serves as an important benchmark for current and future quantum computer performance. Previous benchmarks in this field have focused on variational methods for computing ground and excited states of various molecules, including a benchmarking suite focused on performance of computing ground states for alkali-hydrides under an array of error mitigation methods. Here, we outline state of the art methods to reach chemical accuracy in hybrid quantum-classical electronic structure calculations of alkali hydride molecules on NISQ devices from IBM. We demonstrate how to extend the reach of variational eigensolvers with new symmetry preserving Ans\"atze. Next, we outline how to use quantum imaginary time evolution and Lanczos as a complementary method to variational techniques, highlighting the advantages of each approach. Finally, we demonstrate a new error mitigation method which uses systematic error cancellation via hidden inverse gate constructions, improving the performance of typical variational algorithms. These results show that electronic structure calculations have advanced rapidly, to routine chemical accuracy for simple molecules, from their inception on quantum computers a few short years ago, and they point to further rapid progress to larger molecules as the power of NISQ devices grows. 
% 
% We benchmark hybrid quantum-classical  methods  using  electronic structure calculations on NISQ devices from IBM using alkali hydride test molecules LiH, NaH, KH, RbH. We present a variational eigensolver method focusing on direct minimization using  an  Ansatz  constructed with a symmetry preserving Ansatz and approaches quantum imaginary time evolution and quantum Lanczos method. In addition to these methods we discuss hidden inverses error mitigation strategy with examples that helps reduce the systematic error in NISQ devices. Our results show that...
% 
\end{abstract}

\maketitle

\section{Introduction}
%%\tshc{travis} Background on NISQ computing, role of benchmarks in monitoring progress in quantum computing, remarks on existing benchmarks, eg, gate versus use case
Electronic structure calculation in quantum chemical systems is one of the most important applications of noisy intermediate scale quantum (NISQ) computers. Because of its important role as a key application, it has recently filled the gap in application-level benchmarks of quantum computer performance. Computational chemistry has long played a major role in benchmarking classical computers, starting with interpretation  of molecular spectra against quantum chemistry calculations. Its importance has been recognized in Gerhard Hertzberg's 1971 Nobel lecture awarded ``{\em for his contributions to the knowledge of electronic structure and geometry of molecules}''. Today, a wide range recent electronic structure calculations using variational methods in the literature represent a promising paradigm within which to evaluate NISQ computers as well~\cite{cao2019quantum,mcardle2020quantum,kandala2017,colless,urbanek,Hempel}. Recently these ideas were combined into a benchmark suite that aimed to enumerate the potential of cloud NISQ devices under an array of configurations including active orbitals, Ansatz construction, and error mitigation methods~\cite{McCaskey2019}. Further results in preparing and measuring quantum states for simple molecules indicate a growing capability that may eventually demonstrate quantum computational advantage~\cite{google2020hartree}, which makes tracking and validating performance of NISQ devices in this field all the more important.

% The rapidly evolving progress of noisy intermediate-scale quantum (NISQ) computing underscores the need to test and evaluate new devices and applications. By identifying common measures of both hardware and software, benchmarks are a means to track performance, monitor improvements, and anticipate growth in NISQ capabilities. Like conventional computational benchmarks, NISQ benchmarks must be sensitive to the fundamental performance of device components, such as gates and registers, as well as the means by which computation integrates the components into a meaningful result. 
Notably, recent methods for benchmarking, including quantum volume benchmarking \cite{Blume2020}, cross-entropy benchmarking \cite{Boixo2018}, and randomized benchmarking \cite{Knill2008}, provide good low-level detail about quantum gate performance, but they have lacked an application-specific context that relates to the role of computation as a tool for discovery. This gap is compounded by the fact that performance metrics such as single and two qubit gate fidelities are not always good predictors of higher level application performance. This gap further motivates the use of quantum chemistry as a key performance indicator in the NISQ era. Here, we present the current progress in developing this application area as a performance benchmark, and we outline the accuracy and performance that is achievable for electronic structure calculations on quantum computers today with the state of the art algorithms and error mitigation techniques. We subsequently estimate where progress in this field can move to in the near term in light of these developments.

%% \tshc{travis} Review uses of VQE and QITE methods, role in benchmarking, prior studies and results \cite{Kandala,McCaskey,many others}

\section{Quantum Chemistry Algorithms}
One of the most well-known NISQ algorithms is the variational quantum eigensolver (VQE), which has proven to be especially versatile for testing and tuning applications of quantum chemistry on NISQ devices. In particular, the VQE method offered the first experimental example of an electronic structure calculation by quantum computing on quantum photonic  hardware~\cite{Peruzzo2014}. Upon its first implementation on superconducting hardware, the algorithm was shown to outperform time evolution via Trotterization and subsequent phase estimation~\cite{omalley}. The advantage of variational methods is that low-depth, parameterized circuits produce simple trial wavefunctions, minimizing errors due to decoherence, while classical optimization proved robust to the systematic and depolarization errors found in NISQ devices. Several implementations that followed focused on the utility of unitary coupled cluster methods and variations thereof~\cite{romero,McCaskey2019,grimsley2019adaptive,claudino2020benchmarking,sokolov2020quantum}. Other implementations demonstrated that the trial Ansatz need not be in UCC form to obtain chemical accuracy~\cite{kandala2017}. Eventually, variational methods were modified to be able to compute excited states as well~\cite{colless}.

The key limitation in time-evolution computation outlined in \cite{omalley} motivated a search other techniques to compute time-evolved quantities such as full eigenspectra. 
Quantum imaginary time evolution (QITE) algorithm to find the ground-state energy of many-particle systems is one such method~\cite{McArdle2019}. 
In principle, QITE is capable of preparing exact quantum states without variational optimization. Direct computation of the eigenspectra without ancilla qubits using this algorithm was outlined in \cite{Motta2019}, and the method was further extended to a practical application on NISQ devices in \cite{Yeter2019}. 
The quantum Lanczos method was used in tandem with QITE in these demonstrations in order to recover higher excited states of molecules and nuclei, avoiding the problem of high circuit depth associated with Trotterization and real time evolution. 
Further demonstrations of the QITE algorithm applied to chemistry problems include~\cite{Gomes2020, Barison2020}. These early examples were notably limited to two-qubit computations due to hardware noise. Details on the inner workings of these algorithms were included in a recent review~\cite{Bauer2020}.

\par 
%\tshc{travis} sensitivity to noise
%The versatility of variational methods for quantum chemistry lie in the sensitivity to hardware layout, error rates, and noise. By tuning the parameters of a variational algorithm, a programmed implemented is optimized with respect to a performance metric. This will include accuracy of the computed observable when the expected result is known a priori, as in benchmarking.
%by and significance for quantifying progress in NISQ devices; diversity of strategies for mitigating errors enable more accurate measure of fundamental limits on performance.
\par
Here, we start from this current state of the field and produce new benchmark techniques that extend the performance of both variational methods and QITE/QLanczos algorithms beyond the state of the art in electronic structure calculations on NISQ devices.  
In particular, we apply and obtain experimental data for an Ansatz constructed with a symmetry preserving circuit (SPC) introduced in Refs.~\cite{Barkoutsos2018,Barron2020,Gard2020}. This Ansatz preserves relevant chemical symmetries and has the benefit of minimal parameter count and low depth. Since the main limitation of coupled cluster methods is noise associated with large circuit depth, particularly in the number of CNOTs required, we find that VQE combined with SPCs vastly outperforms coupled cluster methods due to the commensurately lower noise associated with short depth. Further, since these circuits identify target symmetry spaces, they are also capable of probing select excited states for each unique combination of symmetry eigenvalues. We demonstrate this capability by directly calculating eigenspectra for a range of alkali-hydride molecules. Importantly, we show that VQE+SPC enables chemical accuracy in all of the test molecules in our suite over the cloud for the first time. We also compare and contrast the strengths of VQE+SPC with QITE/QLanczos and demonstrate the use of QITE on 4 qubit depths for the first time in chemistry problems. We make extensive use of error mitigation in our benchmarks, and here we also outline a new error mitigation technique via the use of hidden inverse gate constructions and demonstrate its use in variational algorithms. We find that this error mitigation technique substantially reduces the effects of coherent errors in simulation, further extending the reach of both variational and QITE algorithms by mitigating circuit noise.

% \cite{Jensen2020} computes energy levels of Hydrogen molecule for target intervals. \tshc{I think \cite{Jensen2020} is based on phase estimation and only performed in numerical simulation - maybe out of scope here}.

The rest of this report proceeds as follows. In Section~\ref{sec:chem}, we discuss the chemical Hamiltonians of the alkali hydride molecules LiH, NaH, KH, RbH which are used in this paper as a suite of model chemical systems for benchmarking. In Section~\ref{sec:qcomp}, we discuss three hybrid quantum-classical algorithms that are used to calculate the eigenvalues of the molecules of interest on a quantum computer. Specifically, in Section~\ref{sec:spc}, we discuss SPCs in VQE and present our data obtained using IBM's quantum computers. We then discuss QITE and QLanczos algorithms in Section~\ref{sec:QITE} and present our experimental data collected from IBM's quantum hardware. In Section~\ref{sec:hidden}, we introduce and demonstrate the hidden inverse as an error mitigation strategy for variational algorithms and demonstrate its application on a noisy quantum computer simulator for LiH and NaH molecules. Finally, we conclude and present further predictions for the state of the field in Section~\ref{sec:con}.

%%%%%%%%%%%%%%%%%%%%%%%%%%%%%%%%%%%%%%%%%%%
\section{Chemistry Model}
\label{sec:chem}

In this section, we introduce the underlying model of chemistry  used  throughout this paper. We start by discussing electronic structure of  alkali hydrides molecules  LiH, NaH, KH, RbH.  First, we discuss  model  of Hamiltonian in the second quantization and sketch the embedding of the quantum computation kernel in Hartree-Fock  which separates small active set of orbitals from the remaining orbital  treated with a mean-field approach. Our model of the quantum region includes  four spin orbitals and requires four qubits when using the Jordan-Wigner transformation.
Finally,  we discuss  eigenstates that  correspond to our model Hamiltonian. 

The test molecular systems used in this work are alkali hydride molecules  LiH, NaH, KH, RbH which, respectively, have 4, 12, 20,  38 electrons.
The minimal STO-3G basis set is  used and the corresponding Hilbert space is spanned by, respectively, 6, 10, 14, 23 spherical atomic  orbitals and twice as many molecular spin-orbitals.
Even though we use only a minimal basis  set, the size of the Hilbert space is too large  for a direct computation of electronic structure on NISQ devices with the customary Jordan-Wigner transformation, as it requires  that each  spin orbital  is represented by an individual qubit. 
Fortuitously, this problem can be simplified.  In the majority of  chemical processes, such as  bond breaking or formation, only  the  highest energy, valence  electrons play a significant (active) role, whereas  the tightly  bound core electrons are largely (inactive)  spectators  whose role is limited to a mean-field screening of the Coulomb electrostatic field from the nuclei.
The resulting effective Hamiltonian in the second quantization language is given  by \cite{McCaskey2019}: 
\begin{equation}
H =  H_0 + H_1 + H_2,
\label{eq:Htot}
\end{equation}
with $H_0$ given by
\begin{equation}
H_{0} =  E_{nucl}  +\sum_a \big( {h}^a_a + \tfrac{1}{2} \sum_{b} \bar{g}^{ab}_{ab}\big),
\label{eq:H0}
\end{equation}
where  $a$ and $b$  run over inactive occupied  spin-orbitals of frozen-core. The  $E_{nucl}$ describes  Coulomb repulsion  between bare nuclei cores and the second term  describes the effect of screening by the  core electrons. 
Similarly,  the 1-body  term is given by
\begin{equation}
H_{1} = \sum_{p,q}  \hat p^\dagger  \hat q   \cdot
\Big( {h}^{p}_q + \tfrac{1}{2}
\sum_a   \bar{g}^{a p }_{a q} \Big)
\label{eq:H1}
\end{equation}
where the  term ${h}^{p}_q$ represents interaction of valence electrons with all core ions and the second term  in the parenthesis describes its screening by core electrons. Finally, the 2-body part is
\begin{equation}
H_2 = \tfrac{1}{4} \sum_{p,q,r,s}   \bar{g}^{pq}_{sr}  \cdot  \hat p^\dagger \hat q^\dagger \hat r \hat s.
\label{eq:H2}
\end{equation}
where the indices $p$, $q$, $r$, $s$ run over active spin orbitals whereas indices $a$ and $b$ run over inactive spin-orbitals of a frozen core. The symbols  $h^p_q$  and $\bar g^{pq}_{sr}$ denote, respectively, matrix elements of  core Hamiltonian (kinetic energy of electrons plus Coulomb interaction with core ions) and an anti-symmetrized repulsion  integral:
\begin{equation}
\bar g^{pq}_{sr}=  g^{pq}_{sr}- g^{pq}_{r,s}=  \langle p, q| s,r  \rangle - \langle p, q| r,s  \rangle.   
\end{equation}
This Hamiltonian provides a powerful recipe for embedding of the quantum electronic structure calculation into a classically computed environment obtained via Hartree-Fock (HF) \cite{McCaskey2019,HF-embedding}. 

For the model alkali hydride molecules studied here, only the 2 highest energy electrons are involved  in chemical bonding. 
The selected active space includes only the highest occupied  and  lowest unoccupied molecular orbitals (HOMO and LUMO) and four molecular spin orbitals.
The resulting Hilbert space  spanned by  the  HOMO and LUMO is of O($2^4$) size and  can describe up to  four electrons. The neutral  case  is described by placing two electrons  within the active space.  The Hamiltonian matrix has 16$\times$16 elements and  is block diagonal with each block corresponding to a different  number electrons, with  block size of 1, 4, 6, 4, and 1 for, respectively, 0, 1, 2, 3, and 4 electrons cases. In summary, the effective Hamiltonian applied allows for a significant reduction of quantum resources requiring only 4 qubits for the Jordan-Wigner transformed Hamiltonian. This construction is effectively the state of the art in quantum chemistry computations to date. While higher basis sets have been demonstrated, our benchmark basis set and subsequent qubit numbers are well-matched to today's hardware. Subsequent implementations of these techniques underway will use larger basis sets and enable benchmarking of larger numbers of qubits as the coherence time of the larger cloud devices increases.

For the 4-qubit problems we investigate here, there are $2^4=16$ eigenstates which we can group into distinct symmetry subspaces. Denoting the number of electrons within active space as $N_e$  and corresponding spin projection as $s_z$ one can write parametrized vectors associated with each Hamiltonian block as labeled by their distinguishable quantum numbers in general form:
\begin{eqnarray}
\ket{N_e=1,s_z=0.5} &=&\alpha_1\ket{1000}+\alpha_2\ket{0100} \nonumber\\
\ket{N_e=2,s_z=0} ~~&=& \beta_1\ket{1010}+\beta_2\ket{1001} \nonumber\\
&&+\beta_3 \ket{0110}+\beta_4 \ket{0101} \nonumber \\
\ket{N_e=3,s_z=0.5} &=&\gamma_1\ket{1110}+\gamma_2\ket{1101} \, ,
\label{eq:symmetries}
\end{eqnarray}
where we used  spin block notation such that the first two bits in the ket occupation vector correspond to spin-up orbital whereas the last two bits in the  ket correspond to spin-down orbitals. Next, we note that kets corresponding  $s_z=-0.5$  can be easily obtained by symmetry relations. The trivial states for  $N_e=0$ ($\ket{0000}$) and $N_e=4$ ($\ket{1111}$) have been skipped.
 The second formula in Eq. \ref{eq:symmetries} for $N_e=2$ parametrizes three singlet states ($s$=0) and one triplet state ($s$=1) mixed. 
 For the $N_e=2$ case, the singlet states can be written as
\begin{eqnarray}
\ket{N_e=2,s=0,s_z=0}& = & \beta_1' \ket{1010} + \beta_2'  \ket{0101}  \label{eq:singlet}   \\
      & &  + \beta_3' \cdot \frac{1}{\sqrt{2}}(\ket{1001}+\ket{0110}) \nonumber 
%\label{eq:singlet}      
\end{eqnarray}

whereas the triplet states are
\begin{eqnarray}
&\ket{N_e=2,s=1,s_z=1}&=\ket{1100} \nonumber\\
&\ket{N_e=2,s=1,s_z=0}&=\frac{1}{\sqrt{2}}(\ket{1001}-\ket{0110}) \nonumber\\
&\ket{N_e=2,s=1,s_z=-1}&=\ket{0011} 
\label{eq:trip3}\, .
\end{eqnarray}
Note that of these, only the one with $s_z=0$ is an entangled state, the remaining states are separable. The ground state of our model chemistry is a singlet state model described  by Eq. \ref{eq:singlet}.

%%%%%%%%%%%%%%%%%%%%%%%%%%%%%%%%%%%%%%%%%%%%%%%%%%%%%%%%%%%

\section{Quantum  Variational Electronic Structure Computations}
\label{sec:qcomp}

In this section we discuss application of various variational quantum algorithms towards the estimation of ground  and excited state electronic structure for model chemical Hamiltonians discussed  in Sec.~\ref{sec:chem}.  First we explain the symmetry preserving circuits in which trial state vectors are constructed to preserve spin projection during the variational search. As an illustration, these circuits are used to explore ground and excited states of our model systems on IBM Q hardware.
% The accuracy  of  quantum computing algorithms in the NISQ regime  critically depends on the  sensitivity to hardware layout, error rates, and noise. 
Next, we discuss imaginary time evolution and quantum Lanczos eigensolvers and its application to alkali hydrides  on IBM Q hardware.  
Finally, we illustrate the importance of noise mitigation and introduce the hidden inverse method as an approach to reduce noise in quantum computing. 

%%%%%%%%%%%%%%%%%%%%%%%%%%%%%%%%%%%%%%%%%%%%%%%%%%
\subsection{Symmetry Preserving Circuits}
\label{sec:spc}
% Table of symbols used in this section
% A - ASWAP gate
% \theta - parameter of ASWAP gate, sometimes subscripted
% \phi - parameter of ASWAP gate, sometimes subscripted
% n - number of spin-orbitals 
% m - number of electrons
% s_z - spin projection
% N - total number of qubits, N=2n
% \alpha - coefficients for expansion in symmetry subspace, sometimes subscripted
% \beta - coefficients for expansion in symmetry subspace, sometimes subscripted
% R - parameterized gate
% R_z - rotation about Z axis, e^(-i\theta \sigma_z / 2)
% R_y - rotation about Y axis, e^(-i\theta \sigma_y / 2)
% X - X gate
% H - Hadamard gate
In a standard VQE, parameterized Ans\"atze are chosen that, ideally, are accurate, have a small amount of parameters and a low enough circuit depth for NISQ hardware. Symmetry Preserving Circuits (SPC)~\cite{Barkoutsos2018,Gard2020,Barron2020} have very efficient implementation on NISQ hardware, which we demonstrate here on metal alkali hydrides, experimentally for the first time. Since all of the molecules we consider can be mapped using the Jordan-Wigner mapping onto 4 qubits, the methodology of constructing the SPC results in a fairly simple, compact set of circuits. We note that the qubit tapering methods using symmetry information can also be used to improve performance by reducing circuit width~\cite{Setia}. The SPC method is complementary in its focus on circuit depth. Moreover, SPC uses the minimum number of parameters needed to describe arbitrary states with constraints on symmetries.  These circuits are constructed following specific rules to preserve natural symmetries which exist in the problem Hamiltonian and are maintained through the Jordan-Wigner mapping. Specifically, in a system of $2n$ qubits, each qubit represents the occupation of a spin-orbital, and if we assume a block assignment of spins, then the $n$ ``top'' qubits correspond to spin-up orbitals, while the $n$ remaining ``bottom'' qubits correspond to spin-down orbitals. The SPC methodology constructs a circuit with gates that preserve the number of electrons ($N_e$) and spin projection ($s_z$). The primitive gate used in the SPC construction is a parameterized SWAP-type gate (denoted as ASWAP), given in the computational basis by,
\begin{eqnarray}
    A(\theta ,\phi )=\begin{pmatrix}
    1 & 0 & 0 & 0\\
    0 & \cos \theta  & e^{i\phi }\sin \theta  & 0\\
    0 & e^{-i\phi }\sin \theta & -\cos \theta & 0\\
    0 & 0 & 0 &1
    \end{pmatrix}.\label{eq:Agate}
\end{eqnarray}
Since a chemical Hamiltonian is real-valued and symmetric (Hermitian) hence its energy eigenvalues are real (time-reversal symmetry). %, so are its energy eigenvalues. 
Therefore, we can safely set $\phi=0$ in Eq.\ \eqref{eq:Agate}. By inspection, we can also see that Eq.\ \eqref{eq:Agate} naturally preserves particle number as it only acts non-trivially on the subspace spanned by $\{ \ket{01} , \ket{10} \}$ as controlled by the parameter $\theta$. Similarly, ASWAP also preserves spin projection if the gate is not placed in a way that mixes the up-spin and down-spin subspaces, which is achieved by the construction described in Ref.~\cite{Gard2020}. Under the Jordan-Wigner mapping, total spin is a non-local quantity and therefore it is not simple to conserve using local (two-qubit) gates. Consequently, we focus only on preserving the local quantities of particle number and spin projection, using circuits constructed using Eq.\ \eqref{eq:Agate}. 

%For the 4-qubit problems we investigate here, there are $2^4=16$ eigenstates which we can group into distinct symmetry subspaces. It is convenient to choose the subspaces
%\begin{eqnarray}
%\ket{N_e=1,s_z=0.5} &=&\alpha_1\ket{1000}+\alpha_2\ket{0100} \nonumber\\
%\ket{N_e=2,s_z=0} ~~&=& \beta_1\ket{1010}+\beta_2\ket{1001} \nonumber\\
%&&+\beta_3 \ket{0110}+\beta_4 \ket{0101} \nonumber \\
%\ket{N_e=3,s_z=0.5} &=&\gamma_1\ket{1110}+\gamma_2\ket{1101} \, ,
%\label{eq:symmetries}
%\end{eqnarray}
%labeled by their distinguishable quantum numbers. 
%Note that in this case we have suppressed the total spin ($s$) eigenvalue since the SPC does not directly preserve it. 
% The second formula in Eq. \ref{eq:symmetries} for $N_e=2$ parametrizes    three singlet states
%($s$=0) and one triplet state ($s$=1). 
%In principle, each symmetry space corresponds to a distinct circuit constructed of ASWAP gates and  the SPC cannot directly distinguish between states of different total spin.  
%For the $N_e=2$ case, the singlet states can be written as
%\begin{eqnarray}
%\ket{N_e=2,s=0,s_z=0}& = & \beta_1' \ket{1010} + \beta_2'  \ket{0101}   \\
%      & &  + \beta_3' \cdot \frac{1}{\sqrt{2}}(\ket{1001}+\ket{0110}) \nonumber 
%\end{eqnarray}
%and the triplet states are
%\begin{eqnarray}
%&\ket{N_e=2,s=1,s_z=1}&=\ket{1100} \nonumber\\
%&\ket{N_e=2,s=1,s_z=0}&=\frac{1}{\sqrt{2}}(\ket{1001}-\ket{0110}) \nonumber\\
%&\ket{N_e=2,s=1,s_z=-1}&=\ket{0011} \label{eq:trip3}\, .
%\end{eqnarray}
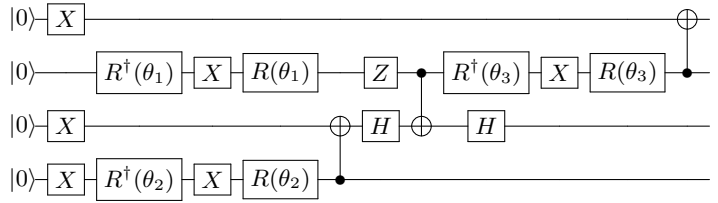
\begin{figure}[!tb]
\[ \Qcircuit @C=0.5em @R=.7em {
\ket{0} &	&	\gate{X}	& \qw	&	\qw	& \qw &	\qw	& \qw  & \qw & \qw &\qw &\qw &	\targ	&	\qw	\\
\ket{0} &	&	\qw	        &	\gate{R^\dagger(\theta_1)}	&	\gate{X}	&	\gate{R(\theta_1)} &\qw	& \gate{Z} & \ctrl{1} & \gate{R^\dagger(\theta_3)}	&	\gate{X}	&	\gate{R(\theta_3)}&	\ctrl{-1}	&	\qw	\\
\ket{0} &	&	\gate{X}	&	\qw	&	\qw & \qw		&	\targ	&	\gate{H} & \targ & \gate{H} &\qw &\qw &\qw &\qw	\\
\ket{0} &	&	\gate{X}	&	\gate{R^\dagger(\theta_2)}	&	\gate{X}	&	\gate{R(\theta_2)}	& \ctrl{-1}& 	\qw	&	\qw &	\qw &	\qw &	\qw &	\qw &	\qw
} \]
\caption{A simplified sample circuit for the case of time-reversal symmetry with $n=4,N_e=2,s_z=0$ spanning the four-dimensional subspace with $N_e=2$ in Eq.~\eqref{eq:symmetries} using a minimal number of parameters (3). Here we define $R(\theta)\equiv R_z(\pi)R_y(\theta+ \frac{\pi}{2})$ with $R_{z}(\phi )= e^{ -i\phi \sigma _{z}/2 }$, $R_{y}(\theta )= e^{ -i\theta \sigma _{y}/2 }$.}
\label{fig:ASWAPcircs1_simp}
\end{figure}
\begin{figure}[!tb]
\[ \Qcircuit @C=1.5em @R=1.2em {
\ket{0} &	&	\gate{X}	&	\qw	        &	\qw &\qw &\targ &\qw	\\
\ket{0} &	&	\qw	        &	\gate{R^\dagger(\theta_1)}        &   \gate{X} & \gate{R(\theta_1)} &\ctrl{-1} & \qw\\
\ket{0} &	&	\qw	        &   \qw	                            &	\qw &	\qw &	\qw &\qw	\\
\ket{0} &	&	\qw       	&	\qw	                            &	\qw &	\qw &	\qw	 &\qw} \]
\caption{A circuit which generates exactly any state with $N_e=1$ in Eq.~\eqref{eq:symmetries}. As explained in the text, by using particle-hole symmetry we obtain a similar circuit for any state with $N_e=3$ in Eq.~\eqref{eq:symmetries}.}
\label{fig:ASWAPcircs2_simp}
\end{figure}
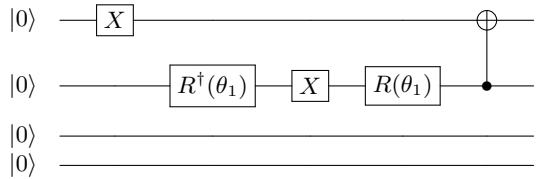

%Note that of these, only the one with $s_z=0$ is an entangled state, the remaining states are %separable. 
In principle SPC are parameterized circuits constructed of ASWAP gates that operate within a small subspace while preserving the number of particles, $N_e$, and spin projection, $s_z$, but not the total spin. 
Moreover, the triplet states in our 4-qubits test case are not parameterized. When we consider full VQE experiments, we therefore are only concerned with generating the singlet state with $s_z=0$.
For completeness, the numerical parameters which generate the triplet state with $s_z=0$ using the circuit in Fig.~\ref{fig:ASWAPcircs1_simp}  are $\theta_1=-\pi/4$, $\theta_2=-\pi/4$, $\theta_3=3\pi/4$.
\par
In Fig.~\ref{fig:ASWAPcircs1_simp} we show the simplified SPC for constructing any state with $N_e=2$ in Eq.~\eqref{eq:symmetries}. For clarity we discuss how to simplify this circuit in the appendix and only present the result here. In total this circuit is composed of 3 CNOT gates, 21 single qubit gates (when expressed in terms of $R_z$, $R_y$, $X$, $H$, though other choices are possible), and the minimal number of parameters for this symmetry subspace, which is 3. 
%We also conjecture that three CNOTs is likely the minimal number of CNOTs possible since the symmetry space in this example spans four basis states and the initial state is a separable state. 
The resulting quantum circuit is very similar to the  recently  proposed UCC-3 circuit in Ref. \cite{McCaskey2019}.
Since the ground state of the molecules we investigate is of the form of a state with $N_e=2$ in Eq.~\eqref{eq:symmetries}, the circuit given by Fig.~\ref{fig:ASWAPcircs1_simp} can efficiently find these ground states.
\par
In Fig.~\ref{fig:ASWAPcircs2_simp}, we similarly show a SPC which finds any state with $N_e=1$ in Eq.~\eqref{eq:symmetries}. Since this symmetry space is smaller, the resulting circuit is simpler. In this case, the total number of CNOT gates is 1, single qubit gates is 6 and the minimal parameter count is 1. Note that the states with $N_e=1$ in Eq.~\eqref{eq:symmetries} are degenerate with those with $N_e=1,s_z=-0.5$. However, we could also build a SPC which finds this degenerate state, since it has a different $s_z$ eigenvalue. We omit finding these degenerate states in practice, but note that they are in principle, distinguishable with SPCs. By particle-hole symmetry we can also view Fig.~\ref{fig:ASWAPcircs2_simp} as the case of a single hole being exchanged (rather than a particle), therefore naturally representing the $N_e=3,s_z=-0.5$ case. In this sense, the $N_e$ and $N-N_e$ particles cases only differ in the initial single qubit gates. Therefore we have two primary configurations of SPC, Fig.~\ref{fig:ASWAPcircs1_simp} and Fig.~\ref{fig:ASWAPcircs2_simp}, for the case of chemical Hamiltonians on 4-qubits.
\par
By employing  direct search of energy minimum and  maximum within each SPC subspace one can find  all of the 16 eigenstates  except one of three singlet states which is neither a minimum nor maximum energy within its own symmetry subspace.
That is, energy minimization (maximization) within the singlet  state subspace leads to the lowest (highest) energy state within this manifold. The third singlet state can be extracted by employing constraint optimization in which the orthogonality to the ground singlet state is enforced.
%.This can be understood by the segmentation of the relatively small space defined by 4-qubits. 
We summarize the total number of states within each symmetry subspace in Table~\ref{tbl:counting}. 
As discussed, the SPC allows to efficiently reduce search space while at the same time employing relatively simple circuits.
\begin{table}[h!]
    \begin{tabular}{|l|c|c|c|}
    \hline
    Symmetry Space &  \# of States &  CNOTs &  Parameters  \\
         \hline
         $N_e=1,s=0.5,s_z=0.5$ & $2$ & $1$ & $1$ \\
         \hline
         $N_e=1,s=0.5,s_z=-0.5$ & $2$ & $1$ & $1$ \\
         \hline
         $N_e=2,s=0,~~s_z=0$ & $3$ & $3$ & $3$ \\
         \hline
         $N_e=2,s=1,~~s_z=0$ & $1$ & $3$ & $0$ \\
         \hline
         $N_e=3,s=0.5,s_z=0.5$ & $2$ & $1$ & $1$ \\
         \hline
         $N_e=3,s=0.5,s_z=-0.5$ & $2$ & $1$ & $1$ \\
         \hline
    \end{tabular}
    \caption{Table of symmetry spaces along with the number of states within each symmetry subspace and the CNOT and parameter count for the corresponding SPC. Note that we omit four trivial, separable states.}
    \label{tbl:counting}
\end{table}
\par
\begin{figure*}
    \centering
    
    \definecolor{mmaBlue}{HTML}{5e81b5}
    \definecolor{mmaOrange}{HTML}{e19c24}
    \definecolor{mmaGreen}{HTML}{8fb032}
    \definecolor{mmaRed}{HTML}{eb6235}
    \definecolor{mmaPurple}{HTML}{8778b3}
    \definecolor{mmaBrown}{HTML}{c56e1a}
    \definecolor{mmaBlack}{HTML}{000000}
    \begin{tikzpicture}
        \begin{groupplot}[
                            group style = {group name=HeH,group size = 2 by 2,
                            x descriptions at=edge bottom,
                            %y descriptions at=edge left,
                            horizontal sep = 0pt,
                            vertical sep = 20pt
                            },
                            width=0.8*\columnwidth,
                            xlabel={Bond Distance (Angstroms)},
                            legend pos=north east,
                            ticklabel style={
                            /pgf/number format/fixed,
                            /pgf/number format/precision=2
                            },legend style={nodes={scale=0.8, transform shape}}
                            ]
        
        \nextgroupplot[ylabel = Energy (Ha),title=LiH]
        
        %plot data
        \addplot[only marks, width=0.75pt,solid,color=mmaBlue,mark=x]
        table[x=distance, y=energy_g, col sep=comma]{LiH_Energy_data.csv};
        \addplot[only marks, width=0.75pt,solid,color=mmaGreen,mark=pentagon] table[x=distance, y=energy_2, col sep=comma]{LiH_Energy_data.csv};
        \addplot[only marks, width=0.75pt,solid,color=mmaRed,mark=o] table[x=distance, y=energy_1, col sep=comma]{LiH_Energy_data.csv};
        \addplot[only marks, width=0.75pt,solid,color=mmaOrange,mark=triangle] table[x=distance, y=energy_3, col sep=comma]{LiH_Energy_data.csv};
        \addplot[only marks, width=0.75pt,solid,color=mmaPurple,mark=square] table[x=distance, y=energy_g_max, col sep=comma]{LiH_Energy_data.csv};
        \addplot[only marks, width=0.75pt,solid,color=mmaBrown,mark=+] table[x=distance, y=energy_1_max, col sep=comma]{LiH_Energy_data.csv};
        \addplot[only marks, width=0.75pt,solid,color=mmaBlack,mark=diamond] table[x=distance, y=energy_3_max, col sep=comma]{LiH_Energy_data.csv};
    
        %plot exact energies
        \addplot[width=0.75pt,solid,color=mmaBlue]
        table[x=distance_e, y=energy_eg, col sep=comma]{LiH_Energy_data.csv};
        \addplot[width=0.75pt,solid,color=mmaGreen]
        table[x=distance_e, y=energy_e2, col sep=comma]{LiH_Energy_data.csv};
        \addplot[width=0.75pt,solid,color=mmaRed]
        table[x=distance_e, y=energy_e1, col sep=comma]{LiH_Energy_data.csv};
        \addplot[width=0.75pt,solid,color=mmaOrange]
        table[x=distance_e, y=energy_e3, col sep=comma]{LiH_Energy_data.csv};
        \addplot[width=0.75pt,solid,color=mmaPurple]
        table[x=distance_e, y=energy_eg_max, col sep=comma]{LiH_Energy_data.csv};
        \addplot[width=0.75pt,solid,color=mmaBrown]
        table[x=distance_e, y=energy_e1_max, col sep=comma]{LiH_Energy_data.csv};
        \addplot[width=0.75pt,solid,color=mmaBlack]
        table[x=distance_e, y=energy_e3_max, col sep=comma]{LiH_Energy_data.csv};
        
        \nextgroupplot[ylabel = Energy (Ha),title=NaH,yticklabel pos= right, ylabel near ticks]
        
        %plot data
        \addplot[only marks, width=0.75pt,solid,color=mmaBlue,mark=x]
        table[x=distance, y=energy_g, col sep=comma]{NaH_Energy_data.csv};
        \addplot[only marks, width=0.75pt,solid,color=mmaGreen,mark=pentagon] table[x=distance, y=energy_2, col sep=comma]{NaH_Energy_data.csv};
        \addplot[only marks, width=0.75pt,solid,color=mmaRed,mark=o] table[x=distance, y=energy_1, col sep=comma]{NaH_Energy_data.csv};
        \addplot[only marks, width=0.75pt,solid,color=mmaOrange,mark=triangle] table[x=distance, y=energy_3, col sep=comma]{NaH_Energy_data.csv};
        \addplot[only marks, width=0.75pt,solid,color=mmaPurple,mark=square] table[x=distance, y=energy_g_max, col sep=comma]{NaH_Energy_data.csv};
        \addplot[only marks, width=0.75pt,solid,color=mmaBrown,mark=+] table[x=distance, y=energy_1_max, col sep=comma]{NaH_Energy_data.csv};
        \addplot[only marks, width=0.75pt,solid,color=mmaBlack,mark=diamond] table[x=distance, y=energy_3_max, col sep=comma]{NaH_Energy_data.csv};
    
        %plot exact energies
        \addplot[width=0.75pt,solid,color=mmaBlue]
        table[x=distance_e, y=energy_eg, col sep=comma]{NaH_Energy_data.csv};
        \addplot[width=0.75pt,solid,color=mmaGreen]
        table[x=distance_e, y=energy_e2, col sep=comma]{NaH_Energy_data.csv};
        \addplot[width=0.75pt,solid,color=mmaRed]
        table[x=distance_e, y=energy_e1, col sep=comma]{NaH_Energy_data.csv};
        \addplot[width=0.75pt,solid,color=mmaOrange]
        table[x=distance_e, y=energy_e3, col sep=comma]{NaH_Energy_data.csv};
        \addplot[width=0.75pt,solid,color=mmaPurple]
        table[x=distance_e, y=energy_eg_max, col sep=comma]{NaH_Energy_data.csv};
        \addplot[width=0.75pt,solid,color=mmaBrown]
        table[x=distance_e, y=energy_e1_max, col sep=comma]{NaH_Energy_data.csv};
        \addplot[width=0.75pt,solid,color=mmaBlack]
        table[x=distance_e, y=energy_e3_max, col sep=comma]{NaH_Energy_data.csv};
        
        \nextgroupplot[ylabel = Energy (Ha),title=KH,yticklabel pos= left, ylabel near ticks]
        \legend{{$N_e=2,s_z=0$},{$N_e=2,s=1,s_z=0$},{$N_e=1,s_z=0.5$},{$N_e=3,s_z=0.5$},{$N_e=2,s_z=0$},{$N_e=1,s_z=0.5$},{$N_e=3,s_z=0.5$}}
        
        %plot data
        \addplot[only marks, width=0.75pt,solid,color=mmaBlue,mark=x]
        table[x=distance, y=energy_g, col sep=comma]{KH_Energy_data.csv};
        \addplot[only marks, width=0.75pt,solid,color=mmaGreen,mark=pentagon] table[x=distance, y=energy_2, col sep=comma]{KH_Energy_data.csv};
        \addplot[only marks, width=0.75pt,solid,color=mmaRed,mark=o] table[x=distance, y=energy_1, col sep=comma]{KH_Energy_data.csv};
        \addplot[only marks, width=0.75pt,solid,color=mmaOrange,mark=triangle] table[x=distance, y=energy_3, col sep=comma]{KH_Energy_data.csv};
        \addplot[only marks, width=0.75pt,solid,color=mmaPurple,mark=square] table[x=distance, y=energy_g_max, col sep=comma]{KH_Energy_data.csv};
        \addplot[only marks, width=0.75pt,solid,color=mmaBrown,mark=+] table[x=distance, y=energy_1_max, col sep=comma]{KH_Energy_data.csv};
        \addplot[only marks, width=0.75pt,solid,color=mmaBlack,mark=diamond] table[x=distance, y=energy_3_max, col sep=comma]{KH_Energy_data.csv};
    
        %plot exact energies
        \addplot[width=0.75pt,solid,color=mmaBlue]
        table[x=distance_e, y=energy_eg, col sep=comma]{KH_Energy_data.csv};
        \addplot[width=0.75pt,solid,color=mmaGreen]
        table[x=distance_e, y=energy_e2, col sep=comma]{KH_Energy_data.csv};
        \addplot[width=0.75pt,solid,color=mmaRed]
        table[x=distance_e, y=energy_e1, col sep=comma]{KH_Energy_data.csv};
        \addplot[width=0.75pt,solid,color=mmaOrange]
        table[x=distance_e, y=energy_e3, col sep=comma]{KH_Energy_data.csv};
        \addplot[width=0.75pt,solid,color=mmaPurple]
        table[x=distance_e, y=energy_eg_max, col sep=comma]{KH_Energy_data.csv};
        \addplot[width=0.75pt,solid,color=mmaBrown]
        table[x=distance_e, y=energy_e1_max, col sep=comma]{KH_Energy_data.csv};
        \addplot[width=0.75pt,solid,color=mmaBlack]
        table[x=distance_e, y=energy_e3_max, col sep=comma]{KH_Energy_data.csv};
        
         \nextgroupplot[ylabel = Energy (Ha),title=RbH,yticklabel pos= right, ylabel near ticks]
        %plot data
        \addplot[only marks, width=0.75pt,solid,color=mmaBlue,mark=x]
        table[x=distance, y=energy_g, col sep=comma]{RbH_Energy_data.csv};
        \addplot[only marks, width=0.75pt,solid,color=mmaGreen,mark=pentagon] table[x=distance, y=energy_2, col sep=comma]{RbH_Energy_data.csv};
        \addplot[only marks, width=0.75pt,solid,color=mmaRed,mark=o] table[x=distance, y=energy_1, col sep=comma]{RbH_Energy_data.csv};
        \addplot[only marks, width=0.75pt,solid,color=mmaOrange,mark=triangle] table[x=distance, y=energy_3, col sep=comma]{RbH_Energy_data.csv};
        \addplot[only marks, width=0.75pt,solid,color=mmaPurple,mark=square] table[x=distance, y=energy_g_max, col sep=comma]{RbH_Energy_data.csv};
        \addplot[only marks, width=0.75pt,solid,color=mmaBrown,mark=+] table[x=distance, y=energy_1_max, col sep=comma]{RbH_Energy_data.csv};
        \addplot[only marks, width=0.75pt,solid,color=mmaBlack,mark=diamond] table[x=distance, y=energy_3_max, col sep=comma]{RbH_Energy_data.csv};
    
        %plot exact energies
        \addplot[width=0.75pt,solid,color=mmaBlue]
        table[x=distance_e, y=energy_eg, col sep=comma]{RbH_Energy_data.csv};
        \addplot[width=0.75pt,solid,color=mmaGreen]
        table[x=distance_e, y=energy_e2, col sep=comma]{RbH_Energy_data.csv};
        \addplot[width=0.75pt,solid,color=mmaRed]
        table[x=distance_e, y=energy_e1, col sep=comma]{RbH_Energy_data.csv};
        \addplot[width=0.75pt,solid,color=mmaOrange]
        table[x=distance_e, y=energy_e3, col sep=comma]{RbH_Energy_data.csv};
        \addplot[width=0.75pt,solid,color=mmaPurple]
        table[x=distance_e, y=energy_eg_max, col sep=comma]{RbH_Energy_data.csv};
        \addplot[width=0.75pt,solid,color=mmaBrown]
        table[x=distance_e, y=energy_e1_max, col sep=comma]{RbH_Energy_data.csv};
        \addplot[width=0.75pt,solid,color=mmaBlack]
        table[x=distance_e, y=energy_e3_max, col sep=comma]{RbH_Energy_data.csv};
        \end{groupplot}
    \end{tikzpicture}
    \caption{Dissociation curves of four molecules run on IBM Hardware (markers) at fixed interatomic distances of $\{0.5,1.0,1.5,2.0,2.5\}$ Angstroms. For each molecule considered, the SPC can accurately probe seven of the eight interesting eigenstates (discussed in text). In all cases, we can see excellent agreement with the exact energies within each symmetry subspace (solid lines). Since the SPC distinguish between different symmetry spaces, we label each curve by its symmetry eigenvalues throughout dissociation.}
    \label{fig:Energies}
\end{figure*}
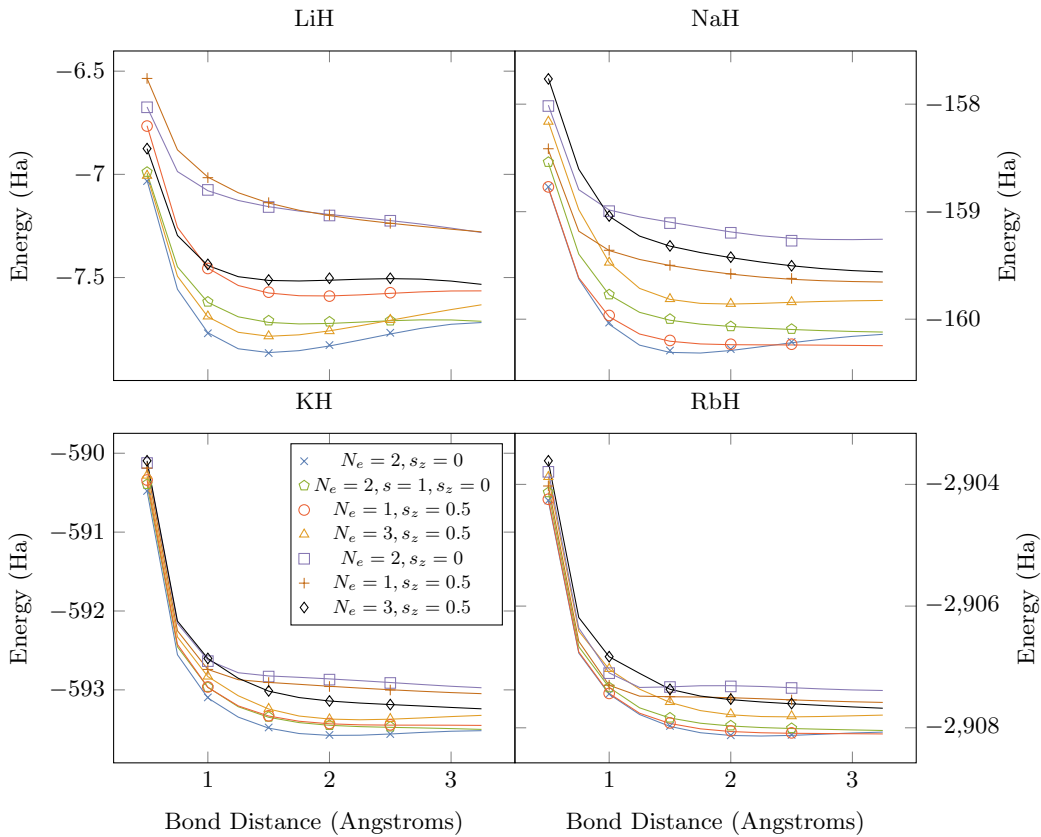

\begin{figure*}
    \centering
    
    \definecolor{mmaBlue}{HTML}{5e81b5}
    \definecolor{mmaOrange}{HTML}{e19c24}
    \definecolor{mmaGreen}{HTML}{8fb032}
    \definecolor{mmaRed}{HTML}{eb6235}
    \definecolor{mmaPurple}{HTML}{8778b3}
    \definecolor{mmaBrown}{HTML}{c56e1a}
    \definecolor{mmaBlack}{HTML}{000000}
    \begin{tikzpicture}
        \begin{groupplot}[
                            group style = {group name=Energy Errors,group size = 2 by 2,
                            x descriptions at=edge bottom,
                            %y descriptions at=edge left,
                            horizontal sep = 0pt,
                            vertical sep = 20pt
                            },
                            width=0.8*\columnwidth,
                            xlabel={Bond Distance (Angstroms)},
                            legend pos=south east,
                            ticklabel style={
                            /pgf/number format/fixed,
                            /pgf/number format/precision=2
                            },legend style={nodes={scale=0.8, transform shape}}
                            ]
        \nextgroupplot[ylabel = Energy Error (Ha),title=LiH,ymode=log,ymin=1e-5,yticklabel pos= left, ylabel near ticks]
        
        %plot data
        \addplot+[
          mmaBlue, mark options={mmaBlue, scale=0.75},
          only marks, mark=x,
          error bars/.cd, 
            y dir=both, 
            y explicit
        ]table[x=distance, y=ene_diff_g,y error=ene_std_g, col sep=comma]{LiH_Energy_data.csv};
        \addplot+[
          mmaGreen, mark options={mmaGreen, scale=0.75},
          only marks, mark=pentagon,
          error bars/.cd, 
            y dir=both, 
            y explicit 
        ]table[x=distance1, y=ene_diff_2,y error=ene_std_2, col sep=comma]{LiH_Energy_data.csv};
        \addplot+[
          mmaRed, mark options={mmaRed, scale=0.75},
          only marks, mark=o,
          error bars/.cd, 
            y dir=both, 
            y explicit 
        ]table[x=distance2, y=ene_diff_1,y error=ene_std_1, col sep=comma]{LiH_Energy_data.csv};
        \addplot+[
          mmaOrange, mark options={mmaOrange, scale=0.75},
          only marks, mark=triangle,
          error bars/.cd, 
            y dir=both, 
            y explicit 
        ]table[x=distance3, y=ene_diff_3,y error=ene_std_3, col sep=comma]{LiH_Energy_data.csv};
        \addplot+[
          mmaPurple, mark options={mmaPurple, scale=0.75},
          only marks, mark=square,
          error bars/.cd, error bar style=solid,
            y dir=both, 
            y explicit 
        ]table[x=distance4, y=ene_diff_g_max,y error=ene_std_g_max, col sep=comma]{LiH_Energy_data.csv};
        \addplot+[
          mmaBrown, mark options={mmaBrown, scale=0.75},
          only marks, mark=+,
          error bars/.cd, error bar style=solid,
            y dir=both, 
            y explicit 
        ]table[x=distance5, y=ene_diff_1_max,y error=ene_std_1_max, col sep=comma]{LiH_Energy_data.csv};
        \addplot+[
          mmaBlack, mark options={mmaBlack, scale=0.75},
          only marks, mark=diamond,
          error bars/.cd, error bar style=solid,
          y fixed,
          y dir=both, 
          y explicit 
        ]table[x=distance6, y=ene_diff_3_max,y error=ene_std_3_max, col sep=comma]{LiH_Energy_data.csv};
        \addplot+[
          mmaBlue, mark=None,
          dashed
        ]table{
        0.3 1.6e-3
        2.9 1.6e-3
        };
        
        \nextgroupplot[ylabel = Energy Error (Ha),title=NaH,ymode=log,ymin=1e-5,yticklabel pos= right, ylabel near ticks]
        
        %plot data
        \addplot+[
          mmaBlue, mark options={mmaBlue, scale=0.75},
          only marks, mark=x,
          error bars/.cd, 
            y dir=both, 
            y explicit
        ]table[x=distance, y=ene_diff_g,y error=ene_std_g, col sep=comma]{NaH_Energy_data.csv};
        \addplot+[
          mmaGreen, mark options={mmaGreen, scale=0.75},
          only marks, mark=pentagon,
          error bars/.cd, 
            y dir=both, 
            y explicit 
        ]table[x=distance1, y=ene_diff_2,y error=ene_std_2, col sep=comma]{NaH_Energy_data.csv};
        \addplot+[
          mmaRed, mark options={mmaRed, scale=0.75},
          only marks, mark=o,
          error bars/.cd, 
            y dir=both, 
            y explicit 
        ]table[x=distance2, y=ene_diff_1,y error=ene_std_1, col sep=comma]{NaH_Energy_data.csv};
        \addplot+[
          mmaOrange, mark options={mmaOrange, scale=0.75},
          only marks, mark=triangle,
          error bars/.cd, 
            y dir=both, 
            y explicit 
        ]table[x=distance3, y=ene_diff_3,y error=ene_std_3, col sep=comma]{NaH_Energy_data.csv};
        \addplot+[
          mmaPurple, mark options={mmaPurple, scale=0.75},
          only marks, mark=square,
          error bars/.cd, error bar style=solid,
            y dir=both, 
            y explicit 
        ]table[x=distance4, y=ene_diff_g_max,y error=ene_std_g_max, col sep=comma]{NaH_Energy_data.csv};
        \addplot+[
          mmaBrown, mark options={mmaBrown, scale=0.75},
          only marks, mark=+,
          error bars/.cd, error bar style=solid,
            y dir=both, 
            y explicit 
        ]table[x=distance5, y=ene_diff_1_max,y error=ene_std_1_max, col sep=comma]{NaH_Energy_data.csv};
        \addplot+[
          mmaBlack, mark options={mmaBlack, scale=0.75},
          only marks, mark=diamond,
          error bars/.cd, error bar style=solid,
          y fixed,
          y dir=both, 
          y explicit 
        ]table[x=distance6, y=ene_diff_3_max,y error=ene_std_3_max, col sep=comma]{NaH_Energy_data.csv};
        \addplot+[
          mmaBlue, mark=None,
          dashed
        ]table{
        0.3 1.6e-3
        2.9 1.6e-3
        };
        
        \nextgroupplot[ylabel = Energy Error (Ha),title=KH,ymode=log,ymin=1e-5,yticklabel pos= left, ylabel near ticks]
        
        %plot data
        \addplot+[
          mmaBlue, mark options={mmaBlue, scale=0.75},
          only marks, mark=x,
          error bars/.cd, 
            y dir=both, 
            y explicit
        ]table[x=distance, y=ene_diff_g,y error=ene_std_g, col sep=comma]{KH_Energy_data.csv};
        \addplot+[
          mmaGreen, mark options={mmaGreen, scale=0.75},
          only marks, mark=pentagon,
          error bars/.cd, 
            y dir=both, 
            y explicit
        ]table[x=distance1, y=ene_diff_2,y error=ene_std_2, col sep=comma]{KH_Energy_data.csv};
        \addplot+[
          mmaRed, mark options={mmaRed, scale=0.75},
          only marks, mark=o,
          error bars/.cd, 
            y dir=both, 
            y explicit 
        ]table[x=distance2, y=ene_diff_1,y error=ene_std_1, col sep=comma]{KH_Energy_data.csv};
        \addplot+[
          mmaOrange, mark options={mmaOrange, scale=0.75},
          only marks, mark=triangle,
          error bars/.cd, 
            y dir=both, 
            y explicit 
        ]table[x=distance3, y=ene_diff_3,y error=ene_std_3, col sep=comma]{KH_Energy_data.csv};
        \addplot+[
          mmaPurple, mark options={mmaPurple, scale=0.75},
          only marks, mark=square,
          error bars/.cd, error bar style=solid,
            y dir=both, 
            y explicit 
        ]table[x=distance4, y=ene_diff_g_max,y error=ene_std_g_max, col sep=comma]{KH_Energy_data.csv};
        \addplot+[
          mmaBrown, mark options={mmaBrown, scale=0.75},
          only marks, mark=+,
          error bars/.cd, error bar style=solid,
            y dir=both, 
            y explicit 
        ]table[x=distance5, y=ene_diff_1_max,y error=ene_std_1_max, col sep=comma]{KH_Energy_data.csv};
        \addplot+[
          mmaBlack, mark options={mmaBlack, scale=0.75},
          only marks, mark=diamond,
          error bars/.cd, error bar style=solid,
          y fixed,
          y dir=both, 
          y explicit 
        ]table[x=distance6, y=ene_diff_3_max,y error=ene_std_3_max, col sep=comma]{KH_Energy_data.csv};
        \addplot+[
          mmaBlue, mark=None,
          dashed
        ]table{
        0.3 1.6e-3
        2.9 1.6e-3
        };
        
        \nextgroupplot[ylabel = Energy Error (Ha),title=RbH,ymode=log,ymin=1e-5,yticklabel pos= right, ylabel near ticks]
        
        %plot data
        \addplot+[
          mmaBlue, mark options={mmaBlue, scale=0.75},
          only marks, mark=x,
          error bars/.cd, 
            y dir=both, 
            y explicit
        ]table[x=distance, y=ene_diff_g,y error=ene_std_g, col sep=comma]{RbH_Energy_data.csv};
        \addplot+[
          mmaGreen, mark options={mmaGreen, scale=0.75},
          only marks, mark=pentagon,
          error bars/.cd, 
            y dir=both, 
            y explicit 
        ]table[x=distance1, y=ene_diff_2,y error=ene_std_2, col sep=comma]{RbH_Energy_data.csv};
        \addplot+[
          mmaRed, mark options={mmaRed, scale=0.75},
          only marks, mark=o,
          error bars/.cd, 
            y dir=both, 
            y explicit 
        ]table[x=distance2, y=ene_diff_1,y error=ene_std_1, col sep=comma]{RbH_Energy_data.csv};
        \addplot+[
          mmaOrange, mark options={mmaOrange, scale=0.75},
          only marks, mark=triangle,
          error bars/.cd, 
            y dir=both, 
            y explicit 
        ]table[x=distance3, y=ene_diff_3,y error=ene_std_3, col sep=comma]{RbH_Energy_data.csv};
        \addplot+[
          mmaPurple, mark options={mmaPurple, scale=0.75},
          only marks, mark=square,
          error bars/.cd, error bar style=solid,
            y dir=both, 
            y explicit 
        ]table[x=distance4, y=ene_diff_g_max,y error=ene_std_g_max, col sep=comma]{RbH_Energy_data.csv};
        \addplot+[
          mmaBrown, mark options={mmaBrown, scale=0.75},
          only marks, mark=+,
          error bars/.cd, error bar style=solid,
            y dir=both, 
            y explicit 
        ]table[x=distance5, y=ene_diff_1_max,y error=ene_std_1_max, col sep=comma]{RbH_Energy_data.csv};
        \addplot+[
          mmaBlack, mark options={mmaBlack, scale=0.75},
          only marks, mark=diamond,
          error bars/.cd, error bar style=solid,
          y fixed,
          y dir=both, 
          y explicit 
        ]table[x=distance6, y=ene_diff_3_max,y error=ene_std_3_max, col sep=comma]{RbH_Energy_data.csv};
        \addplot+[
          mmaBlue, mark=None,
          dashed
        ]table{
        0.3 1.6e-3
        2.8 1.6e-3
        };
        \legend{{$N_e=2,s_z=0$},{$N_e=2,s=1,s_z=0$},{$N_e=1,s_z=0.5$},{$N_e=3,s_z=0.5$},{$N_e=2,s_z=0$},{$N_e=1,s_z=0.5$},{$N_e=3,s_z=0.5$}}

        \end{groupplot}
    \end{tikzpicture}
    \caption{Energy differences from exact eigenenergies of each of the four molecules we consider. Markers are results from IBM hardware with error bars representing one standard deviation. Different symmetry spaces are artificially offset for easier viewing but still have the same fixed interatomic distances shown in Fig.~\ref{fig:Energies}. We show several results within chemical accuracy (dashed line).}
    \label{fig:Energy Error}
\end{figure*}
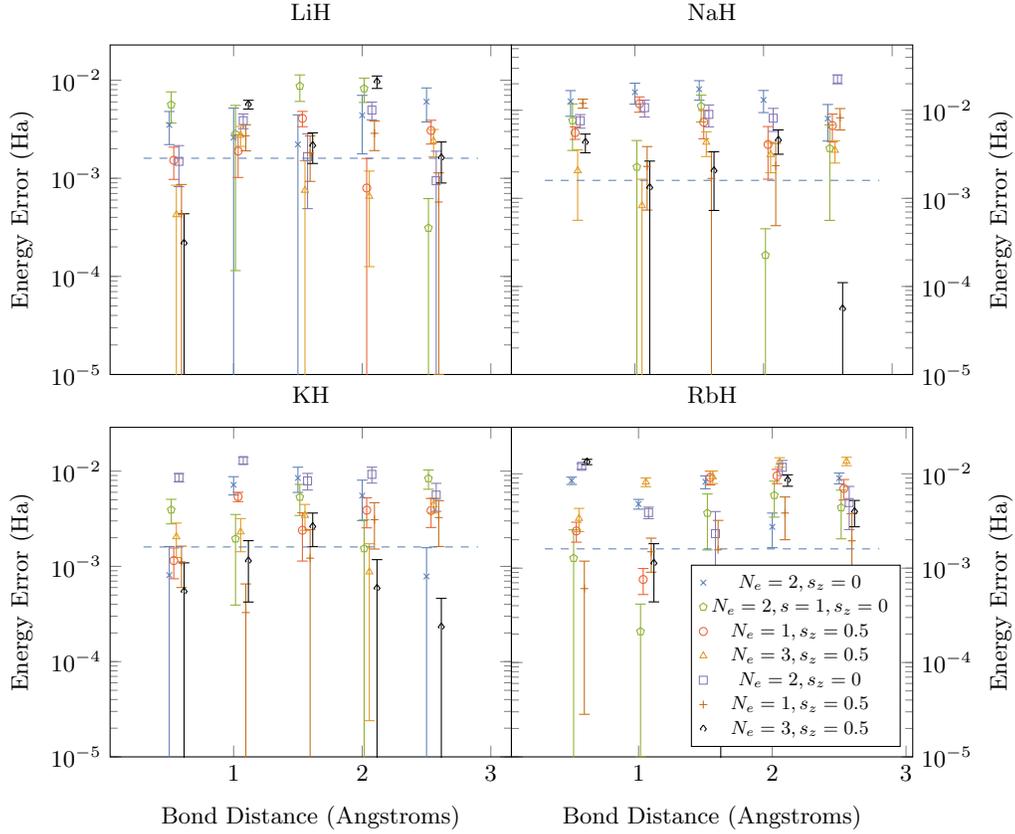
\par
In Fig.~\ref{fig:Energies} we show the dissociation of LiH, NaH, KH, and RbH molecules at five fixed interatomic distances. We plot the exact eigenenergies (found through diagonalization) in solid lines, while results from IBM quantum hardware are shown with markers. Specifically, we ran experiments on IBM Q devices Valencia, Bogota, Santiago, and Vigo but note that results are consistent across these devices. When executing SPC with three CNOT gates (blue, green and purple), we utilize Richardson extrapolation \cite{Li2017} and readout error mitigation strategies. For SPCs with only a single CNOT (red, orange, brown, and black), we chose to only use readout error mitigation as the benefit from Richardson extrapolation is not significant. 
As discussed above  for each symmetry subspace, the SPCs can target two eigenenergies, one as a minimization and one as a maximization optimization within a target subspace. 
%For example, the ground state has symmetry eigenvalues of $N_e=2,s_z=0$ and is found through minimization (blue x's). An excited state with the same symmetry eigenvalues is found through maximization (purple squares). Even when near degeneracies are present, we verify that the correct target state is generated through spin projection and particle number eigenvalues.
\par
In Fig.~\ref{fig:Energy Error} we show the energy difference between the exact target eigenenergy and the energy obtained using IBM hardware. Each marker is offset to reduce crowding but the same fixed interatomic distances are present as shown in Fig.~\ref{fig:Energies}. Error bars are included, representing one standard deviation. We note that many of the standard deviations are of the same order as the energy error mean values. Even though we use the maximum allowed number of shots (8192) this indicates that shot noise is a major source of noise. We can also see that building our SPCs for these three molecules leads to similar results, as well as many measurements within chemical accuracy.

%%%%%%%%%%%%%%%%%%%%%%%%%%%%%%%%%%%%%%%%%%%%%%%%%%%%%%%%%%%%%%%%%%%%%%%%
\subsection{Quantum Imaginary Time Evolution and Quantum Lanczos Algorithms}
\label{sec:QITE}
In this section, we discuss the QITE and QLanczos algorithms which we used to calculate the energy levels for the same suite of molecules as above. The QITE algorithm offers the quantum version of imaginary-time evolution for the non-unitary evolution operator $\mathcal{U}=e^{- \beta H}$. Starting with an appropriately chosen initial state $|\Psi_0\rangle$, after $n$ steps of imaginary-time evolution we obtain
\be
|\Psi(\beta)\rangle=c_n (e^{-\Delta \tau H})^n|\Psi_0\rangle
\ee
where $\Delta \tau=\frac{\beta}{n}$ and $c_n$ is the normalization constant such that $c_n^{-2} =\langle \Psi_0 | \mathcal{U}^2 |\Psi_0 \rangle$. $\beta$ can be thought of as the inverse temperature. As the temperature goes to zero, the system cools down to its ground state as long as the initial state has a non-zero overlap with it. To represent this non-unitary imaginary-time evolution on a quantum computer without the requirement of any ancilla qubits, the QITE algorithm proposes to approximate these non-unitary operators with unitary updates so that the $s$th step of evolution is
\be\label{eq:6}
|\Psi_s\rangle=\frac{c_s}{c_{s-1}}e^{-\Delta \tau H}|\Psi_{s-1}\rangle \approx e^{-i\Delta \tau A[s]}|\Psi_{s-1}\rangle
\ee
with $c_0=1$ and the unitary update operator given in terms of the Hermitian operator
\be
A[s]=\sum_{i_1,\dots,i_{N_q}}a[s]_{i_1,\dots,i_{N_q}} \sigma_{i_1} \dots \sigma_{i_{N_q}}~.\label{eq:As}
\ee
where $N_q$ is the number of qubits and $\sigma \in \{X,Y,Z\}$ are Pauli matrices. Our next step is to calculate the coefficients $a[s]$ by solving Eq.\ \eqref{eq:6} to second order, $\mathcal{O}(\Delta \tau^2)$. This amounts to solving a linear system of equations, $({\bm{\mathcal{S}}}+{\bm{\mathcal{S}}}^T)\cdot {\bm{a}}={\bm{b}}$ where
\be
\mathcal{S}_{\mathcal{I,I'}}=\langle{\sigma_{i_1} \dots \sigma_{i_{N_q}}\sigma_{{i_1'}} \dots \sigma_{{i_{N_q}'}}}\rangle~ \label{eq:Smat}
\ee
and
\begin{equation}
b_{\mathcal{I}}=-i\sqrt{\frac{c_{s-1}}{c_s}}\langle \sigma_{i_1} \dots \sigma_{i_{N_q}} H\rangle \label{eq:bmat}
\end{equation}
with $\mathcal{I}=\{i_1,\dots, i_{N_q}\}$ and similarly for $\mathcal{I}'$. The expectation values at a given step $s$ are calculated with respect to the state in the previous QITE step, $|\Psi_{s-1}\rangle$. The bottleneck in this algorithm is due to fact that the number of measurements needed to calculate ${\bm{\mathcal{S}}}$ (a $3^{N_q}\times 3^{N_q}$ matrix) and ${\bm{b}}$ (a $3^{N_q}$-dimensional vector) scales exponentially with the number of qubits $N_q$ in the system. Having a real Hamiltonian eliminates the terms with an odd number of $Y$ Pauli matrices in ${\bm{b}}$. Also, the fact that the matrix ${\bm{\mathcal{S}}}+{\bm{\mathcal{S}}}^T$ is symmetric reduces the required number of measurements significantly. Nevertheless, the problem of complexity persists, and reducing QITE complexity is an active field of research, just as it has been for reducing circuit depth in VQE (see, e.g., \cite{Tan2020}).

\begin{figure*}
    \centering
    
    \definecolor{mmaBlue}{HTML}{5e81b5}
    \definecolor{mmaOrange}{HTML}{e19c24}
    \definecolor{mmaGreen}{HTML}{8fb032}
    \definecolor{mmaRed}{HTML}{eb6235}
    \definecolor{mmaPurple}{HTML}{8778b3}
    \definecolor{mmaBrown}{HTML}{c56e1a}
    \definecolor{mmaBlack}{HTML}{000000}
    \begin{tikzpicture}
        \begin{groupplot}[
                            group style = {group name=HeH,group size = 2 by 2,
                            x descriptions at=edge bottom,
                            %y descriptions at=edge left,
                            horizontal sep = 0pt,
                            vertical sep = 20pt
                            },
                            width=0.8*\columnwidth,
                            xlabel={Bond Distance (Angstroms)},
                            legend pos=north east,
                            ticklabel style={
                            /pgf/number format/fixed,
                            /pgf/number format/precision=2
                            },legend style={nodes={scale=0.8, transform shape}}
                            ]
        
        \nextgroupplot[ylabel = Energy (Ha),title=LiH]
        
        %plot data
        \addplot[only marks, width=0.75pt,solid,color=mmaBlue,mark=x]
        table[x=distance, y=energy_g, col sep=comma]{LiH_Energy_2qubit_QITE_datacsv.csv};
        \addplot[only marks, width=0.75pt,solid,color=mmaGreen,mark=pentagon] table[x=distance, y=energy_2, col sep=comma]{LiH_Energy_2qubit_QITE_datacsv.csv};
        \addplot[only marks, width=0.75pt,solid,color=mmaRed,mark=o] table[x=distance, y=energy_1, col sep=comma]{LiH_Energy_2qubit_QITE_datacsv.csv};
        \addplot[only marks, width=0.75pt,solid,color=mmaOrange,mark=triangle] table[x=distance, y=energy_3, col sep=comma]{LiH_Energy_2qubit_QITE_datacsv.csv};
        \addplot[only marks, width=0.75pt,solid,color=mmaPurple,mark=square] table[x=distance, y=energy_g_max, col sep=comma]{LiH_Energy_2qubit_QITE_datacsv.csv};
        \addplot[only marks, width=0.75pt,solid,color=mmaBrown,mark=+] table[x=distance, y=energy_1_max, col sep=comma]{LiH_Energy_2qubit_QITE_datacsv.csv};
        \addplot[only marks, width=0.75pt,solid,color=mmaBlack,mark=diamond] table[x=distance, y=energy_3_max, col sep=comma]{LiH_Energy_2qubit_QITE_datacsv.csv};
    
        %plot exact energies
        \addplot[width=0.75pt,solid,color=mmaBlue]
        table[x=distance_e, y=energy_eg, col sep=comma]{LiH_Energy_2qubit_QITE_datacsv.csv};
        \addplot[width=0.75pt,solid,color=mmaGreen]
        table[x=distance_e, y=energy_e2, col sep=comma]{LiH_Energy_2qubit_QITE_datacsv.csv};
        \addplot[width=0.75pt,solid,color=mmaRed]
        table[x=distance_e, y=energy_e1, col sep=comma]{LiH_Energy_2qubit_QITE_datacsv.csv};
        \addplot[width=0.75pt,solid,color=mmaOrange]
        table[x=distance_e, y=energy_e3, col sep=comma]{LiH_Energy_2qubit_QITE_datacsv.csv};
        \addplot[width=0.75pt,solid,color=mmaPurple]
        table[x=distance_e, y=energy_eg_max, col sep=comma]{LiH_Energy_2qubit_QITE_datacsv.csv};
        \addplot[width=0.75pt,solid,color=mmaBrown]
        table[x=distance_e, y=energy_e1_max, col sep=comma]{LiH_Energy_2qubit_QITE_datacsv.csv};
        \addplot[width=0.75pt,solid,color=mmaBlack]
        table[x=distance_e, y=energy_e3_max, col sep=comma]{LiH_Energy_2qubit_QITE_datacsv.csv};
        
        \nextgroupplot[ylabel = Energy (Ha),title=NaH,yticklabel pos= right, ylabel near ticks]
        
        %plot data
        \addplot[only marks, width=0.75pt,solid,color=mmaBlue,mark=x]
        table[x=distance, y=energy_g, col sep=comma]{NaH_Energy_2qubit_QITE_datacsv.csv};
        \addplot[only marks, width=0.75pt,solid,color=mmaGreen,mark=pentagon] table[x=distance, y=energy_2, col sep=comma]{NaH_Energy_2qubit_QITE_datacsv.csv};
        \addplot[only marks, width=0.75pt,solid,color=mmaRed,mark=o] table[x=distance, y=energy_1, col sep=comma]{NaH_Energy_2qubit_QITE_datacsv.csv};
        \addplot[only marks, width=0.75pt,solid,color=mmaOrange,mark=triangle] table[x=distance, y=energy_3, col sep=comma]{NaH_Energy_2qubit_QITE_datacsv.csv};
        \addplot[only marks, width=0.75pt,solid,color=mmaPurple,mark=square] table[x=distance, y=energy_g_max, col sep=comma]{NaH_Energy_2qubit_QITE_datacsv.csv};
        \addplot[only marks, width=0.75pt,solid,color=mmaBrown,mark=+] table[x=distance, y=energy_1_max, col sep=comma]{NaH_Energy_2qubit_QITE_datacsv.csv};
        \addplot[only marks, width=0.75pt,solid,color=mmaBlack,mark=diamond] table[x=distance, y=energy_3_max, col sep=comma]{NaH_Energy_2qubit_QITE_datacsv.csv};
    
        %plot exact energies
        \addplot[width=0.75pt,solid,color=mmaBlue]
        table[x=distance_e, y=energy_eg, col sep=comma]{NaH_Energy_2qubit_QITE_datacsv.csv};
        \addplot[width=0.75pt,solid,color=mmaGreen]
        table[x=distance_e, y=energy_e2, col sep=comma]{NaH_Energy_2qubit_QITE_datacsv.csv};
        \addplot[width=0.75pt,solid,color=mmaRed]
        table[x=distance_e, y=energy_e1, col sep=comma]{NaH_Energy_2qubit_QITE_datacsv.csv};
        \addplot[width=0.75pt,solid,color=mmaOrange]
        table[x=distance_e, y=energy_e3, col sep=comma]{NaH_Energy_2qubit_QITE_datacsv.csv};
        \addplot[width=0.75pt,solid,color=mmaPurple]
        table[x=distance_e, y=energy_eg_max, col sep=comma]{NaH_Energy_2qubit_QITE_datacsv.csv};
        \addplot[width=0.75pt,solid,color=mmaBrown]
        table[x=distance_e, y=energy_e1_max, col sep=comma]{NaH_Energy_2qubit_QITE_datacsv.csv};
        \addplot[width=0.75pt,solid,color=mmaBlack]
        table[x=distance_e, y=energy_e3_max, col sep=comma]{NaH_Energy_2qubit_QITE_datacsv.csv};
        
        \nextgroupplot[ylabel = Energy (Ha),title=KH,yticklabel pos= left, ylabel near ticks]
        \legend{{$N_e=2,s_z=0$},{$N_e=2,s=1,s_z=0$},{$N_e=1,s_z=0.5$},{$N_e=3,s_z=0.5$},{$N_e=2,s_z=0$},{$N_e=1,s_z=0.5$},{$N_e=3,s_z=0.5$}}
        
        %plot data
        \addplot[only marks, width=0.75pt,solid,color=mmaBlue,mark=x]
        table[x=distance, y=energy_g, col sep=comma]{KH_Energy_2qubit_QITE_datacsv.csv};
        \addplot[only marks, width=0.75pt,solid,color=mmaGreen,mark=pentagon] table[x=distance, y=energy_2, col sep=comma]{KH_Energy_2qubit_QITE_datacsv.csv};
        \addplot[only marks, width=0.75pt,solid,color=mmaRed,mark=o] table[x=distance, y=energy_1, col sep=comma]{KH_Energy_2qubit_QITE_datacsv.csv};
        \addplot[only marks, width=0.75pt,solid,color=mmaOrange,mark=triangle] table[x=distance, y=energy_3, col sep=comma]{KH_Energy_2qubit_QITE_datacsv.csv};
        \addplot[only marks, width=0.75pt,solid,color=mmaPurple,mark=square] table[x=distance, y=energy_g_max, col sep=comma]{KH_Energy_2qubit_QITE_datacsv.csv};
        \addplot[only marks, width=0.75pt,solid,color=mmaBrown,mark=+] table[x=distance, y=energy_1_max, col sep=comma]{KH_Energy_2qubit_QITE_datacsv.csv};
        \addplot[only marks, width=0.75pt,solid,color=mmaBlack,mark=diamond] table[x=distance, y=energy_3_max, col sep=comma]{KH_Energy_2qubit_QITE_datacsv.csv};
    
        %plot exact energies
        \addplot[width=0.75pt,solid,color=mmaBlue]
        table[x=distance_e, y=energy_eg, col sep=comma]{KH_Energy_2qubit_QITE_datacsv.csv};
        \addplot[width=0.75pt,solid,color=mmaGreen]
        table[x=distance_e, y=energy_e2, col sep=comma]{KH_Energy_2qubit_QITE_datacsv.csv};
        \addplot[width=0.75pt,solid,color=mmaRed]
        table[x=distance_e, y=energy_e1, col sep=comma]{KH_Energy_2qubit_QITE_datacsv.csv};
        \addplot[width=0.75pt,solid,color=mmaOrange]
        table[x=distance_e, y=energy_e3, col sep=comma]{KH_Energy_2qubit_QITE_datacsv.csv};
        \addplot[width=0.75pt,solid,color=mmaPurple]
        table[x=distance_e, y=energy_eg_max, col sep=comma]{KH_Energy_2qubit_QITE_datacsv.csv};
        \addplot[width=0.75pt,solid,color=mmaBrown]
        table[x=distance_e, y=energy_e1_max, col sep=comma]{KH_Energy_2qubit_QITE_datacsv.csv};
        \addplot[width=0.75pt,solid,color=mmaBlack]
        table[x=distance_e, y=energy_e3_max, col sep=comma]{KH_Energy_2qubit_QITE_datacsv.csv};
        
         \nextgroupplot[ylabel = Energy (Ha),title=RbH,yticklabel pos= right, ylabel near ticks]
        %plot data
        \addplot[only marks, width=0.75pt,solid,color=mmaBlue,mark=x]
        table[x=distance, y=energy_g, col sep=comma]{RbH_Energy_2qubit_QITE_datacsv.csv};
        \addplot[only marks, width=0.75pt,solid,color=mmaGreen,mark=pentagon] table[x=distance, y=energy_2, col sep=comma]{RbH_Energy_2qubit_QITE_datacsv.csv};
        \addplot[only marks, width=0.75pt,solid,color=mmaRed,mark=o] table[x=distance, y=energy_1, col sep=comma]{RbH_Energy_2qubit_QITE_datacsv.csv};
        \addplot[only marks, width=0.75pt,solid,color=mmaOrange,mark=triangle] table[x=distance, y=energy_3, col sep=comma]{RbH_Energy_2qubit_QITE_datacsv.csv};
        \addplot[only marks, width=0.75pt,solid,color=mmaPurple,mark=square] table[x=distance, y=energy_g_max, col sep=comma]{RbH_Energy_2qubit_QITE_datacsv.csv};
        \addplot[only marks, width=0.75pt,solid,color=mmaBrown,mark=+] table[x=distance, y=energy_1_max, col sep=comma]{RbH_Energy_2qubit_QITE_datacsv.csv};
        \addplot[only marks, width=0.75pt,solid,color=mmaBlack,mark=diamond] table[x=distance, y=energy_3_max, col sep=comma]{RbH_Energy_2qubit_QITE_datacsv.csv};
    
        %plot exact energies
        \addplot[width=0.75pt,solid,color=mmaBlue]
        table[x=distance_e, y=energy_eg, col sep=comma]{RbH_Energy_2qubit_QITE_datacsv.csv};
        \addplot[width=0.75pt,solid,color=mmaGreen]
        table[x=distance_e, y=energy_e2, col sep=comma]{RbH_Energy_2qubit_QITE_datacsv.csv};
        \addplot[width=0.75pt,solid,color=mmaRed]
        table[x=distance_e, y=energy_e1, col sep=comma]{RbH_Energy_2qubit_QITE_datacsv.csv};
        \addplot[width=0.75pt,solid,color=mmaOrange]
        table[x=distance_e, y=energy_e3, col sep=comma]{RbH_Energy_2qubit_QITE_datacsv.csv};
        \addplot[width=0.75pt,solid,color=mmaPurple]
        table[x=distance_e, y=energy_eg_max, col sep=comma]{RbH_Energy_2qubit_QITE_datacsv.csv};
        \addplot[width=0.75pt,solid,color=mmaBrown]
        table[x=distance_e, y=energy_e1_max, col sep=comma]{RbH_Energy_2qubit_QITE_datacsv.csv};
        \addplot[width=0.75pt,solid,color=mmaBlack]
        table[x=distance_e, y=energy_e3_max, col sep=comma]{RbH_Energy_2qubit_QITE_datacsv.csv};
        \end{groupplot}
    \end{tikzpicture}
    \caption{Energy vs. bond distance obtained using the reduced Hamiltonian blocks for molecules LiH, KH, RbH, NaH. The experimental data, notated with markers in the figure, were obtained using QITE and QLanczos algorithms. The quantum circuits were run on IBM Q's several quantum computers such as Casablanca, Manhattan, Vigo, Bogota, Rome. The experimental results are in very good agreement with the exact values shown with straight lines in figure. The error bars represent $\pm \sigma$.}
    \label{fig:QITE_energy}
\end{figure*}
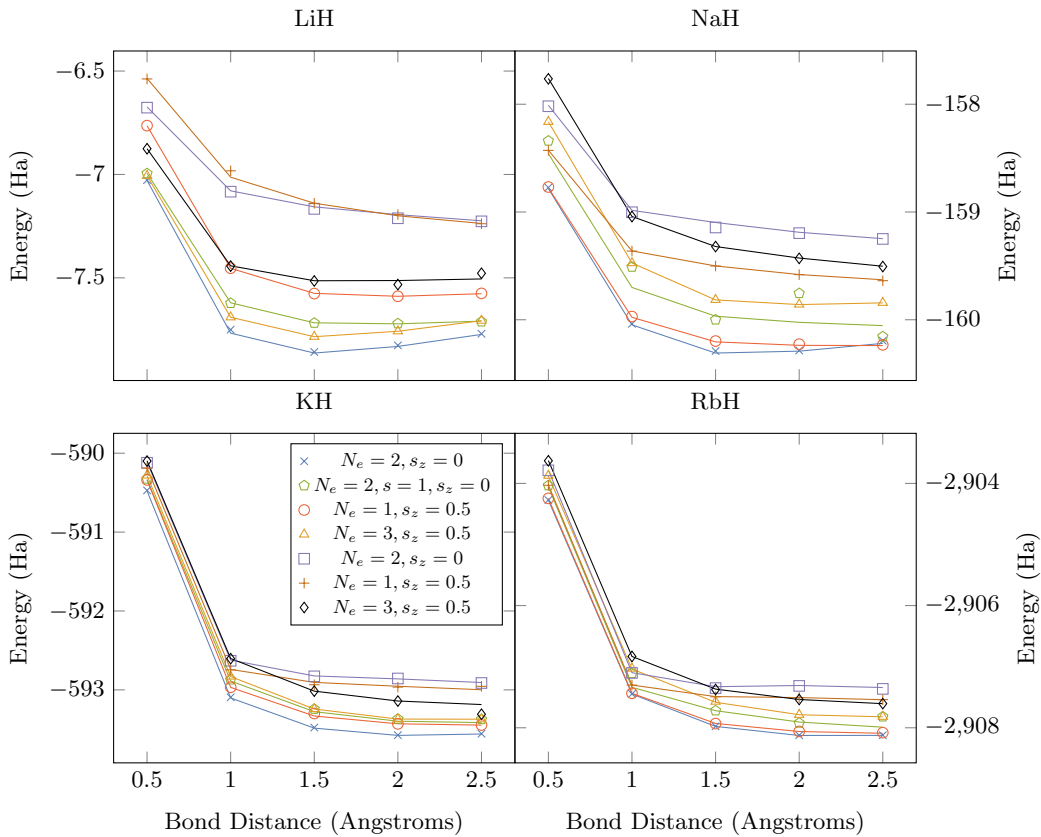

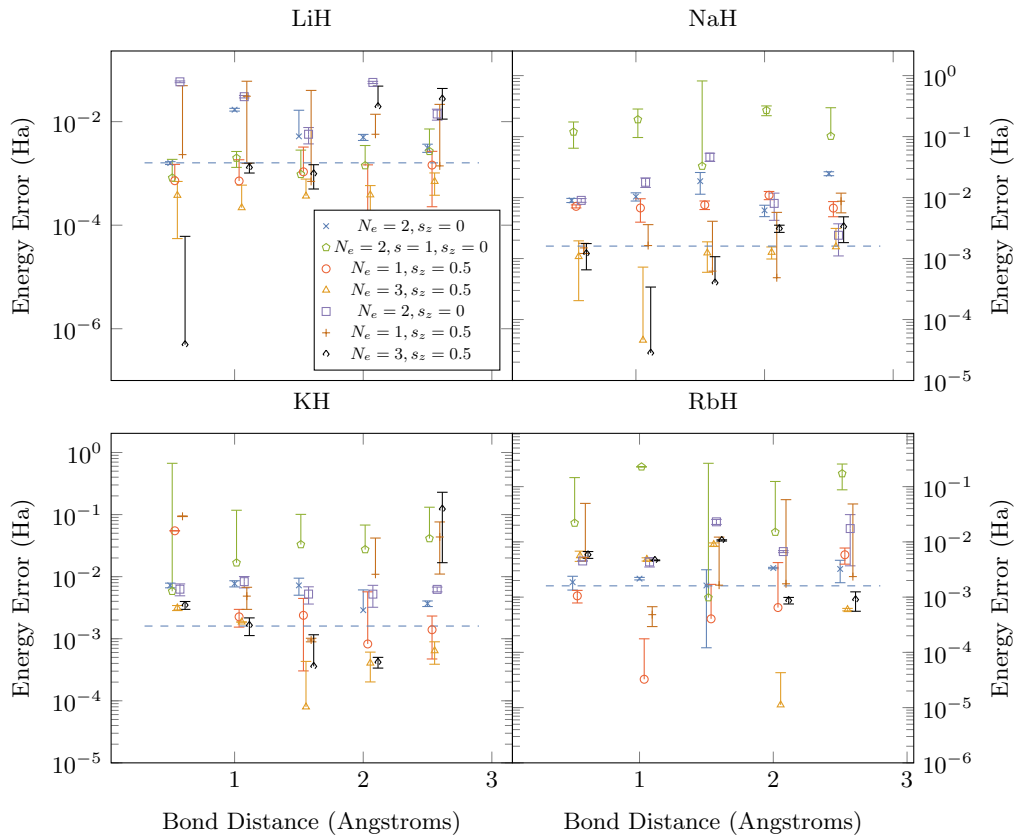
\begin{figure*}
    \centering
    
    \definecolor{mmaBlue}{HTML}{5e81b5}
    \definecolor{mmaOrange}{HTML}{e19c24}
    \definecolor{mmaGreen}{HTML}{8fb032}
    \definecolor{mmaRed}{HTML}{eb6235}
    \definecolor{mmaPurple}{HTML}{8778b3}
    \definecolor{mmaBrown}{HTML}{c56e1a}
    \definecolor{mmaBlack}{HTML}{000000}
    \begin{tikzpicture}
        \begin{groupplot}[
                            group style = {group name=Energy Errors,group size = 2 by 2,
                            x descriptions at=edge bottom,
                            %y descriptions at=edge left,
                            horizontal sep = 0pt,
                            vertical sep = 20pt
                            },
                            width=0.8*\columnwidth,
                            xlabel={Bond Distance (Angstroms)},
                            legend pos=south east,
                            ticklabel style={
                            /pgf/number format/fixed,
                            /pgf/number format/precision=2
                            },legend style={nodes={scale=0.7, transform shape}}
                            ]
        \nextgroupplot[ylabel = Energy Error (Ha),title=LiH,ymode=log,ymin=1e-7,yticklabel pos= left, ylabel near ticks]
        \legend{{$N_e=2,s_z=0$},{$N_e=2,s=1,s_z=0$},{$N_e=1,s_z=0.5$},{$N_e=3,s_z=0.5$},{$N_e=2,s_z=0$},{$N_e=1,s_z=0.5$},{$N_e=3,s_z=0.5$}}
        %plot data
        \addplot+[
          mmaBlue, mark options={mmaBlue, scale=0.75},
          only marks, mark=x,
          error bars/.cd, 
            y dir=both, 
            y explicit
        ]table[x=distance, y=ene_diff_g,y error=ene_std_g, col sep=comma]{LiH_Energy_2qubit_QITE_datacsv.csv};
        \addplot+[
          mmaGreen, mark options={mmaGreen, scale=0.75},
          only marks, mark=pentagon,
          error bars/.cd, 
            y dir=both, 
            y explicit 
        ]table[x=distance1, y=ene_diff_2,y error=ene_std_2, col sep=comma]{LiH_Energy_2qubit_QITE_datacsv.csv};
        \addplot+[
          mmaRed, mark options={mmaRed, scale=0.75},
          only marks, mark=o,
          error bars/.cd, 
            y dir=both, 
            y explicit 
        ]table[x=distance2, y=ene_diff_1,y error=ene_std_1, col sep=comma]{LiH_Energy_2qubit_QITE_datacsv.csv};
        \addplot+[
          mmaOrange, mark options={mmaOrange, scale=0.75},
          only marks, mark=triangle,
          error bars/.cd, 
            y dir=both, 
            y explicit 
        ]table[x=distance3, y=ene_diff_3,y error=ene_std_3, col sep=comma]{LiH_Energy_2qubit_QITE_datacsv.csv};
        \addplot+[
          mmaPurple, mark options={mmaPurple, scale=0.75},
          only marks, mark=square,
          error bars/.cd, error bar style=solid,
            y dir=both, 
            y explicit 
        ]table[x=distance4, y=ene_diff_g_max,y error=ene_std_g_max, col sep=comma]{LiH_Energy_2qubit_QITE_datacsv.csv};
        \addplot+[
          mmaBrown, mark options={mmaBrown, scale=0.75},
          only marks, mark=+,
          error bars/.cd, error bar style=solid,
            y dir=both, 
            y explicit 
        ]table[x=distance5, y=ene_diff_1_max,y error=ene_std_1_max, col sep=comma]{LiH_Energy_2qubit_QITE_datacsv.csv};
        \addplot+[
          mmaBlack, mark options={mmaBlack, scale=0.75},
          only marks, mark=diamond,
          error bars/.cd, error bar style=solid,
          y fixed,
          y dir=both, 
          y explicit 
        ]table[x=distance6, y=ene_diff_3_max,y error=ene_std_3_max, col sep=comma]{LiH_Energy_2qubit_QITE_datacsv.csv};
        \addplot+[
          mmaBlue, mark=None,
          dashed
        ]table{
        0.3 1.6e-3
        2.9 1.6e-3
        };
        
        \nextgroupplot[ylabel = Energy Error (Ha),title=NaH,ymode=log,ymin=1e-5,yticklabel pos= right, ylabel near ticks]
        % \legend{{$N_e=2,s_z=0$},{$N_e=2,s=1,s_z=0$},{$N_e=1,s_z=0.5$},{$N_e=3,s_z=0.5$},{$N_e=2,s_z=0$},{$N_e=1,s_z=0.5$},{$N_e=3,s_z=0.5$}}
        %plot data
        \addplot+[
          mmaBlue, mark options={mmaBlue, scale=0.75},
          only marks, mark=x,
          error bars/.cd, 
            y dir=both, 
            y explicit
        ]table[x=distance, y=ene_diff_g,y error=ene_std_g, col sep=comma]{NaH_Energy_2qubit_QITE_datacsv.csv};
        \addplot+[
          mmaGreen, mark options={mmaGreen, scale=0.75},
          only marks, mark=pentagon,
          error bars/.cd, 
            y dir=both, 
            y explicit 
        ]table[x=distance1, y=ene_diff_2,y error=ene_std_2, col sep=comma]{NaH_Energy_2qubit_QITE_datacsv.csv};
        \addplot+[
          mmaRed, mark options={mmaRed, scale=0.75},
          only marks, mark=o,
          error bars/.cd, 
            y dir=both, 
            y explicit 
        ]table[x=distance2, y=ene_diff_1,y error=ene_std_1, col sep=comma]{NaH_Energy_2qubit_QITE_datacsv.csv};
        \addplot+[
          mmaOrange, mark options={mmaOrange, scale=0.75},
          only marks, mark=triangle,
          error bars/.cd, 
            y dir=both, 
            y explicit 
        ]table[x=distance3, y=ene_diff_3,y error=ene_std_3, col sep=comma]{NaH_Energy_2qubit_QITE_datacsv.csv};
        \addplot+[
          mmaPurple, mark options={mmaPurple, scale=0.75},
          only marks, mark=square,
          error bars/.cd, error bar style=solid,
            y dir=both, 
            y explicit 
        ]table[x=distance4, y=ene_diff_g_max,y error=ene_std_g_max, col sep=comma]{NaH_Energy_2qubit_QITE_datacsv.csv};
        \addplot+[
          mmaBrown, mark options={mmaBrown, scale=0.75},
          only marks, mark=+,
          error bars/.cd, error bar style=solid,
            y dir=both, 
            y explicit 
        ]table[x=distance5, y=ene_diff_1_max,y error=ene_std_1_max, col sep=comma]{NaH_Energy_2qubit_QITE_datacsv.csv};
        \addplot+[
          mmaBlack, mark options={mmaBlack, scale=0.75},
          only marks, mark=diamond,
          error bars/.cd, error bar style=solid,
          y fixed,
          y dir=both, 
          y explicit 
        ]table[x=distance6, y=ene_diff_3_max,y error=ene_std_3_max, col sep=comma]{NaH_Energy_2qubit_QITE_datacsv.csv};
        \addplot+[
          mmaBlue, mark=None,
          dashed
        ]table{
        0.3 1.6e-3
        2.9 1.6e-3
        };
        
        \nextgroupplot[ylabel = Energy Error (Ha),title=KH,ymode=log,ymin=1e-5,yticklabel pos= left, ylabel near ticks]
        
        %plot data
        \addplot+[
          mmaBlue, mark options={mmaBlue, scale=0.75},
          only marks, mark=x,
          error bars/.cd, 
            y dir=both, 
            y explicit
        ]table[x=distance, y=ene_diff_g,y error=ene_std_g, col sep=comma]{KH_Energy_2qubit_QITE_datacsv.csv};
        \addplot+[
          mmaGreen, mark options={mmaGreen, scale=0.75},
          only marks, mark=pentagon,
          error bars/.cd, 
            y dir=both, 
            y explicit
        ]table[x=distance1, y=ene_diff_2,y error=ene_std_2, col sep=comma]{KH_Energy_2qubit_QITE_datacsv.csv};
        \addplot+[
          mmaRed, mark options={mmaRed, scale=0.75},
          only marks, mark=o,
          error bars/.cd, 
            y dir=both, 
            y explicit 
        ]table[x=distance2, y=ene_diff_1,y error=ene_std_1, col sep=comma]{KH_Energy_2qubit_QITE_datacsv.csv};
        \addplot+[
          mmaOrange, mark options={mmaOrange, scale=0.75},
          only marks, mark=triangle,
          error bars/.cd, 
            y dir=both, 
            y explicit 
        ]table[x=distance3, y=ene_diff_3,y error=ene_std_3, col sep=comma]{KH_Energy_2qubit_QITE_datacsv.csv};
        \addplot+[
          mmaPurple, mark options={mmaPurple, scale=0.75},
          only marks, mark=square,
          error bars/.cd, error bar style=solid,
            y dir=both, 
            y explicit 
        ]table[x=distance4, y=ene_diff_g_max,y error=ene_std_g_max, col sep=comma]{KH_Energy_2qubit_QITE_datacsv.csv};
        \addplot+[
          mmaBrown, mark options={mmaBrown, scale=0.75},
          only marks, mark=+,
          error bars/.cd, error bar style=solid,
            y dir=both, 
            y explicit 
        ]table[x=distance5, y=ene_diff_1_max,y error=ene_std_1_max, col sep=comma]{KH_Energy_2qubit_QITE_datacsv.csv};
        \addplot+[
          mmaBlack, mark options={mmaBlack, scale=0.75},
          only marks, mark=diamond,
          error bars/.cd, error bar style=solid,
          y fixed,
          y dir=both, 
          y explicit 
        ]table[x=distance6, y=ene_diff_3_max,y error=ene_std_3_max, col sep=comma]{KH_Energy_2qubit_QITE_datacsv.csv};
        \addplot+[
          mmaBlue, mark=None,
          dashed
        ]table{
        0.3 1.6e-3
        2.9 1.6e-3
        };
        
        \nextgroupplot[ylabel = Energy Error (Ha),title=RbH,ymode=log,ymin=1e-6,yticklabel pos= right, ylabel near ticks]
        
        %plot data
        \addplot+[
          mmaBlue, mark options={mmaBlue, scale=0.75},
          only marks, mark=x,
          error bars/.cd, 
            y dir=both, 
            y explicit
        ]table[x=distance, y=ene_diff_g,y error=ene_std_g, col sep=comma]{RbH_Energy_2qubit_QITE_datacsv.csv};
        \addplot+[
          mmaGreen, mark options={mmaGreen, scale=0.75},
          only marks, mark=pentagon,
          error bars/.cd, 
            y dir=both, 
            y explicit 
        ]table[x=distance1, y=ene_diff_2,y error=ene_std_2, col sep=comma]{RbH_Energy_2qubit_QITE_datacsv.csv};
        \addplot+[
          mmaRed, mark options={mmaRed, scale=0.75},
          only marks, mark=o,
          error bars/.cd, 
            y dir=both, 
            y explicit 
        ]table[x=distance2, y=ene_diff_1,y error=ene_std_1, col sep=comma]{RbH_Energy_2qubit_QITE_datacsv.csv};
        \addplot+[
          mmaOrange, mark options={mmaOrange, scale=0.75},
          only marks, mark=triangle,
          error bars/.cd, 
            y dir=both, 
            y explicit 
        ]table[x=distance3, y=ene_diff_3,y error=ene_std_3, col sep=comma]{RbH_Energy_2qubit_QITE_datacsv.csv};
        \addplot+[
          mmaPurple, mark options={mmaPurple, scale=0.75},
          only marks, mark=square,
          error bars/.cd, error bar style=solid,
            y dir=both, 
            y explicit 
        ]table[x=distance4, y=ene_diff_g_max,y error=ene_std_g_max, col sep=comma]{RbH_Energy_2qubit_QITE_datacsv.csv};
        \addplot+[
          mmaBrown, mark options={mmaBrown, scale=0.75},
          only marks, mark=+,
          error bars/.cd, error bar style=solid,
            y dir=both, 
            y explicit 
        ]table[x=distance5, y=ene_diff_1_max,y error=ene_std_1_max, col sep=comma]{RbH_Energy_2qubit_QITE_datacsv.csv};
        \addplot+[
          mmaBlack, mark options={mmaBlack, scale=0.75},
          only marks, mark=diamond,
          error bars/.cd, error bar style=solid,
          y fixed,
          y dir=both, 
          y explicit 
        ]table[x=distance6, y=ene_diff_3_max,y error=ene_std_3_max, col sep=comma]{RbH_Energy_2qubit_QITE_datacsv.csv};
        \addplot+[
          mmaBlue, mark=None,
          dashed
        ]table{
        0.3 1.6e-3
        2.8 1.6e-3
        };
        % \legend{{$N_e=2,s_z=0$},{$N_e=2,s=1,s_z=0$},{$N_e=1,s_z=0.5$},{$N_e=3,s_z=0.5$},{$N_e=2,s_z=0$},{$N_e=1,s_z=0.5$},{$N_e=3,s_z=0.5$}}

        \end{groupplot}
    \end{tikzpicture}
    \caption{Error in energy vs. bond distance obtained using the reduced Hamiltonian blocks for molecules LiH, KH, RbH, NaH. The dashed line in each panel represents the chemical accuracy ($1.5 \times 10^{-3}$). As seen in the figures we were able to obtain several energy eigenvalues within the chemical accuracy.}
    \label{fig:energy_error}
\end{figure*}

\begin{figure}
    \centering
    
    \definecolor{mmaBlue}{HTML}{5e81b5}
    \definecolor{mmaOrange}{HTML}{e19c24}
    \definecolor{mmaGreen}{HTML}{8fb032}
    \definecolor{mmaRed}{HTML}{eb6235}
    \definecolor{mmaPurple}{HTML}{8778b3}
    \definecolor{mmaBrown}{HTML}{c56e1a}
    \definecolor{mmaBlack}{HTML}{000000}
    \begin{tikzpicture}
        \begin{axis}[
            xlabel={Bond Distance (Angstroms)},
            ylabel={Energy (Ha)},
            ticklabel style={
                /pgf/number format/fixed,
                /pgf/number format/precision=2
            },
            legend style={nodes={scale=0.8, transform shape}}
        ]
        %\begin{groupplot}[
        %                    group style = {group name=HeH,group size = 2 by 2,
        %                    x descriptions at=edge bottom,
        %                    %y descriptions at=edge left,
        %                    horizontal sep = 0pt,
        %                    vertical sep = 20pt
        %                    },
        %                    width=0.8*\columnwidth,
        %                    xlabel={Bond Distance (Angstroms)},
        %                    legend pos=north east,
        %                    ticklabel style={
        %                    /pgf/number format/fixed,
        %                    /pgf/number format/precision=2
        %                    },legend style={nodes={scale=0.5, transform shape}}
        %                    ]
        
        %\nextgroupplot[ylabel = Energy (Ha),title=LiH]
        %\nextgroupplot[ylabel = Energy (Ha),title=LiH,yticklabel pos= left, ylabel near ticks]
        
        %plot data
        \addplot[only marks, width=0.75pt,solid,color=mmaBlue,mark=x,mark size=3pt]
        table[x=distance, y=energy_g, col sep=comma]{LiH_Energy_Fourqubit_QITE_data.csv};
        \addplot[only marks, width=0.75pt,solid,color=mmaGreen,mark=pentagon,mark size=4pt]
        table[x=distance, y=energy_2, col sep=comma]{LiH_Energy_Fourqubit_QITE_data.csv};
        \addplot[only marks, width=0.75pt,solid,color=mmaRed,mark=o,mark size=3pt]
        table[x=distance, y=energy_1, col sep=comma]{LiH_Energy_Fourqubit_QITE_data.csv};
        \addplot[only marks, width=0.75pt,solid,color=mmaOrange,mark=triangle,mark size=3pt]
        table[x=distance, y=energy_3, col sep=comma]{LiH_Energy_Fourqubit_QITE_data.csv};
        \addplot[only marks, width=0.75pt,solid,color=mmaPurple,mark=square,mark size=3pt]
        table[x=distance, y=energy_g_max, col sep=comma]{LiH_Energy_Fourqubit_QITE_data.csv};
        \addplot[only marks, width=0.75pt,solid,color=mmaBrown,mark=+,mark size=3pt]
        table[x=distance, y=energy_1_max, col sep=comma]{LiH_Energy_Fourqubit_QITE_data.csv};
        \addplot[only marks, width=0.75pt,solid,color=mmaBlack,mark=diamond,mark size=3pt]
        table[x=distance, y=energy_3_max, col sep=comma]{LiH_Energy_Fourqubit_QITE_data.csv};
    
        %plot exact energies
        \addplot[width=0.75pt,solid,color=mmaBlue]
        table[x=distance_e, y=energy_eg, col sep=comma]{LiH_Energy_Fourqubit_QITE_data.csv};
        \addplot[width=0.75pt,solid,color=mmaGreen]
        table[x=distance_e, y=energy_e2, col sep=comma]{LiH_Energy_Fourqubit_QITE_data.csv};
        \addplot[width=0.75pt,solid,color=mmaRed]
        table[x=distance_e, y=energy_e1, col sep=comma]{LiH_Energy_Fourqubit_QITE_data.csv};
        \addplot[width=0.75pt,solid,color=mmaOrange]
        table[x=distance_e, y=energy_e3, col sep=comma]{LiH_Energy_Fourqubit_QITE_data.csv};
        \addplot[width=0.75pt,solid,color=mmaPurple]
        table[x=distance_e, y=energy_eg_max, col sep=comma]{LiH_Energy_Fourqubit_QITE_data.csv};
        \addplot[width=0.75pt,solid,color=mmaBrown]
        table[x=distance_e, y=energy_e1_max, col sep=comma]{LiH_Energy_Fourqubit_QITE_data.csv};
        \addplot[width=0.75pt,solid,color=mmaBlack]
        table[x=distance_e, y=energy_e3_max, col sep=comma]{LiH_Energy_Fourqubit_QITE_data.csv};
        \legend{{$N_e=2,s_z=0$},{$N_e=2,s=1~s_z=0$},{$N_e=1,s_z=0.5$},{$N_e=3,s_z=0.5$},{$N_e=2,s_z=0$},{$N_e=1,s_z=0.5$},{$N_e=3,s_z=0.5$}}
        \end{axis}
        %\end{groupplot}
    \end{tikzpicture}
    \caption{Energy vs. bond distance obtained using the 4-qubit LiH molecule Hamiltonian. The Richardson extrapolation applied experimental data, notated with markers in the figure, were obtained using QITE algorithm. The quantum circuits were run on IBM Q's Yorktown and Melbourne device. The experimental results are in agreement with the exact values shown with straight lines in figure but most of them are not in chemical accuracy. The error bars represent $\pm \sigma$.}
    \label{fig:4qubitQITE_energy}
\end{figure}
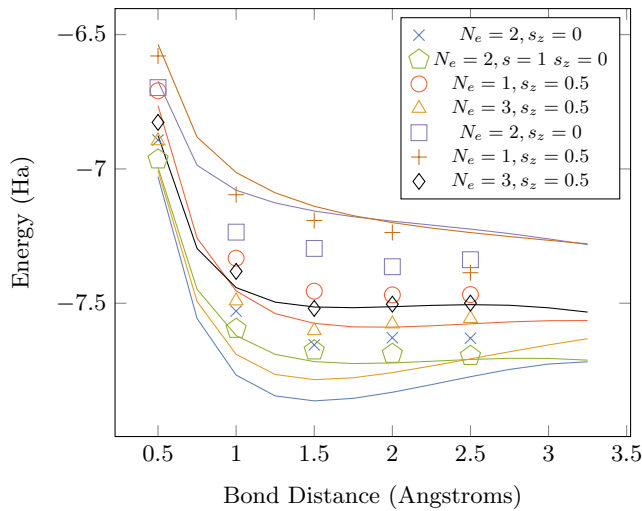

The cloud access to NISQ devices limits the number of measurements that can be done in a reasonable amount of time. Additionally, the depth of quantum circuits is limited by the decoherence time of the qubits. Because of these limitations, we perform QITE algorithms on a NISQ device by following these steps:
\begin{itemize}
    \item To decide on the convergence of the energy expectation value, we use exact QITE calculations. This allows us to validate the hardware results even though the noise in quantum hardware might result in deviations of energy values that are greater than the value that we choose for our convergence criterion. We set the latter to $\epsilon=0.001$ for energy expectation values, and find the number of steps required for convergence.
    \item We then calculate the unitary updates in \eqref{eq:As} at every QITE step from \eqref{eq:Smat} and \eqref{eq:bmat} by performing the energy expectation value calculations on a noisy quantum simulator which includes the noise model of the quantum hardware to be used.
    \item After choosing the initial state $|\Psi_0\rangle$ using the symmetry considerations for each of the reduced Hamiltonian blocks, we find the state 
    \[
    |\Psi_s\rangle=e^{-i\Delta \tau A[s]}e^{-i\Delta \tau A[s-1]}\cdot \cdot \cdot e^{-i\Delta \tau A[1]}|\Psi_0\rangle
    \]
    after application of $s$ unitary updates.
    \item We obtain the quantum circuit corresponding to the state $|\Psi_s\rangle$ by using the \textit{isometry} function in IBM Qiskit library. This function is based on \cite{Iten2020} which decomposes a given state into single-qubit and Controlled-NOT (CNOT) gates with the aim of having the least number of CNOT gates. An example of this quantum circuit can be seen in Fig.~\ref{fig:2qubitcirc}. This quantum circuit has the same single-qubit $U3$ and two-qubit CNOT gates at every QITE step and only the angles of $U3$ gates are updated at every QITE step, where 
    \be
    U3(\theta,\phi,\lambda)=
    \begin{pmatrix} \cos{\frac{\theta}{2}} && -e^{i\lambda} \sin{\frac{\theta}{2}} \\ 
    e^{i\phi}\sin{\frac{\theta}{2}} && e^{i(\phi+\lambda)}\cos{\frac{\theta}{2}}
    \end{pmatrix}~.
    \ee
    \item We then run the quantum circuits corresponding to the states $|\Psi_s\rangle$ on quantum hardware and obtain the energy expectation values at every QITE step. (The experimental results of these calculations can be seen in Fig.~\ref{fig:QITE_energy}).
\end{itemize}
Next, we discuss how to implement the QLanczos algorithm.

The QLanczos algorithm makes use of the QITE algorithm to obtain the states in Krylov space $\mathcal{K}$ which is spanned by $\{ |\Phi_0\rangle, |\Phi_2\rangle, \dots \}$, where $|\Phi_l\rangle \in \{ |\Psi_0\rangle, |\Psi_2\rangle, \dots \}$. After choosing the vectors in Krylov space, we calculate the overlap ($\mathcal{T}$) and Hamiltonian ($\mathcal{H}$) matrices from the QITE algorithm measurements on quantum hardware. The normalization constants are calculated recursively using
\be
\frac{1}{c_{r+1}^2}=\frac{\langle \Phi_r | e^{-2 \Delta \tau H}|\Phi_r\rangle}{c_r^2} \label{cr}
\ee
Since we are using a small $\Delta \tau$ approximation, we calculate the expectation value by expanding to first order in $\Delta \tau$, $\langle \Phi_r | e^{-2 \Delta \tau H}|\Phi_r\rangle = 1-2 \Delta \tau \langle \Phi_r | H|\Phi_r\rangle
% +2\Delta \tau^2 \langle \Phi_r | H^2|\Phi_r\rangle
+\mathcal{O}(\Delta \tau^2)$. Thus, we obtain the normalization constants using the energy expectation values obtained from quantum hardware in the implementation of the QITE algorithm. To increase the accuracy of the calculations, one can use higher-order approximations.

The overlap and Hamiltonian matrix elements can be expressed in terms of these constants and energy expectation values,
\be
\mathcal{T}_{l,l'}=\langle \Phi_l|\Phi_{l'}\rangle=\frac{c_l c_{l'}}{c_r^2}~,
\ee
and 
\be
\mathcal{H}_{l,l'}=\langle \Phi_l|H|\Phi_{l'}\rangle=\mathcal{T}_{l,l'}\langle \Phi_r|H|\Phi_{r}\rangle~,
\ee
where $r=\frac{l+l'}{2}$, and $l, l'$ are even integers. 

The energy eigenvalues $E$ are found by solving the generalized eigenvalue equation 
\be
{\bm{\mathcal{H}x}}=E {\bm{\mathcal{T}x}}~, \label{eq:geneig}
\ee
The corresponding eigenvectors $\bm{x} = (x_0,x_1 ,\dots)$ determine the eigenstates of the system Hamiltonian,
\be
|\Psi[E]\rangle=c_E\left(x_0|\Phi_0\rangle+x_1|\Phi_2\rangle+\dots \right)
\ee
where $c_E^{-1} = \| \sum_{l} x_l |\Phi_l\rangle \|$.
% is then 
% \be
% E = \langle \Psi[E] |H|\Psi[E] \rangle~. \label{QLancEn}
% \ee
% These approximations are easily expressed in terms of quantities that were deduced from QITE. We obtain
% \be E = \frac{\sum_{l,l'=0,2,\dots} x_l^{(E)} \mathcal{H}_{l,l'} x_{l'}^{(E)}}{\sum_{l,l'=0,2,\dots} x_l^{(E)} \mathcal{T}_{l,l'} x_{l'}^{(E)}} \ee 

As explained in the previous section, the symmetry of each of these molecules leads to a simplification of the system which makes them good candidates for implementation on NISQ devices. In the previous section we used this symmetry to simplify the quantum circuits. As a second perspective, we present a simplification of the system by reducing the Hamiltonian into smaller blocks so that we can study these molecules using a smaller number of qubits on quantum hardware. For benchmarking purposes, we also present our QITE algorithm results for the full LiH molecule Hamiltonian for comparison with the results presented using the SPC method. In what follows, we explain the reduction of the system size using the symmetry of the Hamiltonian of various molecules. 

%In Table~\ref{tbl:counting}, there are four states that are exactly known without further calculation. The eight remaining states  can be found by separating the block diagonal terms in each 4-qubit molecular Hamiltonian after omitting the two degenerate states with $N_e=1, s=0.5, s_z=-0.5$ and those with $N_e=3, s=0.5, s_z=-0.5$.
In Section \ref{sec:chem} we discuss chemical  Hamiltonian for our benchmark molecules.  Some of the solutions are given in Eqs. \ref{eq:symmetries},  \ref{eq:singlet} and \ref{eq:trip3},  and also listed in Table~\ref{tbl:counting}.
In a 4-qubit molecular Hamiltonian, the matrix elements corresponding to a no-electron state in the active space ($N_e=0, s_z=0$) and a 4-electron state ($N_e=4, s_z=0$) are diagonal so no further computation of eigenvalues and corresponding eigenstates is needed. The Hamiltonian blocks formed by single ($N_e=1$) and three ($N_e=3$) electron states are  $4\times4$  matrices. 
However, each $4\times4$ Hamiltonian block is itself block-diagonal consisting of two $2\times2$  blocks corresponding to a different value of spin projection ($s_z=+{1\over 2}$ and $s_z=-{1\over 2}$). Each of the four $2\times2$ blocks can be handled by a single-qubit circuit. Similarly, two-electron states ($N_e=2$) form a $6\times 6$ matrix which includes 
a diagonal $4\times 4$ Hamiltonian block  for $s_z=0$ (three singlet states and one triplet state) and 
can be represented using a 2-qubit circuit. The remaining  two states represent triplet states with $s_z=1+$  and $s_z=-1$.
 In conclusion, we can compute the energy levels of each $2^4 \times 2^4$ molecular Hamiltonian using only two-qubit and single-qubit circuits.
 %with some of the eigenvalues deduced exactly directly from the Hamiltonian without any additional calculation. 
% and can be extended to an $8\times 8$ matrix so that it can be studied using a 3-qubit system. 
% In general, $N$-neutrino system can be studied in mass basis using maximum $N-1$ qubits with $2^{N-1}-{N \choose k}$ extra eigenvalues substituted to match the system size with correct number of qubits, where $k$ is the number of particles that give the maximum value for ${N \choose k}$.
This  analysis shows that decomposition of the full Hamiltonian into diagonal blocks based on the number of electron and spin symmetries is essential  making quantum computation compatible with NISQ devices.

We start by calculating the energy expectation values of the system using the QITE algorithm which will give us the ground state energies of the symmetry sector that the initial state belongs to. To this end, as mentioned earlier, the Hamiltonian of the molecules are divided into smaller blocks and the initial states are chosen accordingly using symmetry which results in a smaller number of imaginary-time steps required for convergence. To be able to represent these reduced Hamiltonian blocks on a quantum computer we need to express them in terms of Pauli matrices. This can be done by writing the block Hamiltonian $H_{\text{block}}$ as 
\be H_{\text{block}} =\sum_I c_{I} \sigma_I \ee 
where $I= \{ i_0 , \dots, i_{n_q -1} \}$ and $\sigma_I = \sigma_{i_0}\sigma_{i_1}\dots\sigma_{n_{q}-1}$, with $\sigma \in \{I,X,Y,Z\}$ and $n_q$ the number of qubits used for this block Hamiltonian.
%
%\be
%\begin{split}
%H_{\text{block}}&=\sum_{i_0,i_1,\dots,i_{N_{q'}-1}}\, c_{\sigma_{i_0}\sigma_{i_1}\dots\sigma_{N_{q'}-1}} \sigma_{i_0}\sigma_{i_1}\dots\sigma_{N_{q'}-1}
%\\&=\sum_I c_{\sigma_I} \sigma_I
%\end{split}
%\ee
% where $\sigma_I=\bigotimes_{p=0}^{N_{q'}-1}\sigma_{i_p}$ and $I$ denotes $i_0,i_1,...,i_{N_{q'}-1}$
%with $\sigma_{i_p}=\underbrace{\mathbb{I}\otimes\dots }_{p}\otimes \sigma_{i}\underbrace{\otimes\dots \otimes \mathbb{I}}_{N_{q'}-p-1}$ and $\sigma \in \{I,X,Y,Z\}$. 
The coefficients $c_{I}$ are found from
%calculated using the following equation
\be
c_{I}= \frac{1}{2^{n_{q}}}\text{Tr} [H_{\text{block}} \cdot \sigma_I]
\ee
%where $N_{q'}$ is the number of qubits that is obtained depending on the size of the $H_{\text{block}}$ Hamiltonian. 
Once the coefficients $c_{I}$ are determined, the Hamiltonian $H_{\text{block}}$ expressed in terms of Pauli matrices can be implemented on a quantum computer. 
% For example, for 4-neutrino system there are 6 2-particle states which form a block that can be written as
% \begin{eqnarray}
% H_{\text{block}} &=& \frac{(1 + 6 \mu)}{4} I+H_1+H_2\label{3qubitreducedHam}
% \end{eqnarray}
% with 
% \begin{equation}
%     \begin{split}
%         H_1=\frac{\mu}{2} \big(&X_0\cdot X_1\cdot X_2 +  Z_0\cdot X_1\cdot X_2 + Y_0\cdot Y_1\cdot X_2 \\&+ X_0\cdot X_1 +  X_0\cdot Z_1 + Z_0\cdot X_1 +Y_0\cdot Y_1 \\&+X_1\cdot X_2 + Z_1\cdot X_2 + X_0 + X_1 + X_2 \big)~,
%     \end{split}
% \end{equation}
% \begin{equation}
%     \begin{split}
%         H_2=&\frac{(2 - \mu)}{2} Z_0 \cdot Z_1  + \frac{1}{4} Z_1\cdot Z_2  + \frac{(1 + \mu)}{2}Z_0 \\&+  \frac{(-1 + 2 \mu)}{4} Z_1 + \frac{1}{4} Z_2~,
%     \end{split}
% \end{equation}
% where we added ones to the last two elements of the diagonal to match the matrix size with a 3-qubit system therefore, $N_{q'}=3$. 
We use $H_{\text{block}}$ in QITE and QLanczos algorithms to calculate the energy eigenvalues of each molecular Hamiltonian with states shown in Table ~\ref{tbl:counting}.

We use the information from symmetry considerations of the reduced block Hamiltonians to choose different initial states, $|\Psi_0\rangle$, so as we obtain all energy levels of the system Hamiltonian.

\subsubsection*{QITE and QLanczos Results}

To calculate the energy expectation values at each QITE step for the molecules that we study, we first ran the quantum circuits obtained using the \textit{isometry} function for each QITE step on IBM's cloud accessible quantum computers. We did this for the reduced Hamiltonian blocks first and then for the 4-qubit full LiH molecule Hamiltonian. An example of 2-qubit quantum circuit for reduced block Hamiltonian of $N_e=2$ states can be found in Fig.~\ref{fig:2qubitcirc}. At each QITE step the angles $\theta_1, \theta_2$ and $\theta_3$ change but the quantum circuit depth remains fixed. We then used the measured energy expectation values to find the energy eigenvalues of the systems of interest. 

% \begin{figure}
%     \centering
%     \includegraphics[scale=1]{2Qubit11IsometryCircuit.pdf}
%     \caption{The quantum circuit of each QITE step for 2-qubit reduced Hamiltonian blocks. At every QITE step the angles $\theta_1$, $\theta_2$ and $\theta_3$ change but the circuit depth is fixed.   }
%     \label{fig:2qubitcirc}
% \end{figure}

The calculation of the energy eigenvalue from the QLanczos algorithm using the method described above is often numerically unstable due to the noise introduced by NISQ devices. To stabilize the results, we worked as follows. We only used two-dimensional Krylov spaces for the ${\bm{\mathcal{T}}}$ and ${\bm{\mathcal{H}}}$ matrices in \eqref{eq:geneig}, which have been shown to yield more accurate results in NISQ devices \cite{YeterAydeniz2020}. The Hamiltonian matrices are diagonally dominant with the diagonal elements being a large negative number since due to the electron--nuclei Coulomb interaction. Each reduced Hamiltonian block was shifted  by appropriate  constant value, to ensure that all matrix elements are of the same order. While this did not affect the eigenstates of the system, it improved the accuracy of the results significantly. The accuracy of the eigenvalues can be improved further if one includes higher-order terms in the expansion of $e^{-\Delta\tau H}$, but this would require more measurements of higher moments of the Hamiltonian resulting in longer runs on quantum hardware.

% \begin{figure*}
%     \centering
%     \includegraphics[scale=0.45]{QITEFigures.png}
%     \caption{The convergence of the energy expectation values as a function of imaginary-time, $\beta$, using QITE algorithm. (a) Convergence to the ground state of $N_e=2$ symmetry subspace for LiH molecule with initial state $|\Psi_0\rangle=|11\rangle$ and $-H$. The converged energy corresponds to the negative of the maximum (3rd excited state) energy of that symmetry subspace.  (b) Convergence to the ground state of $N_e=2$ symmetry subspace for RbH molecule with initial state $|\Psi_0\rangle=|00\rangle$ and $H$. The readout error mitigated experimental data were collected on IBM Q 7-qubit Casablanca device and compared to the data calculated using exact diagonalization. Error bars represent $\pm \sigma$.}
%     \label{fig:QITE}
% \end{figure*}

See Appendix~\ref{sec:QITE_app} for an example application of QITE algorithm on IBM Q hardware for 2-qubit reduced Hamiltonian blocks of molecules LiH and RbH where we also demonstrate the 2-qubit circuit obtained from \textit{isometry} function of IBM's Qiskit library.
% In Fig.~\ref{fig:QITE} we present two examples of the measured energy expectation values as a function of imaginary time obtained from the QITE algorithm. Since the QITE algorithm always converges to the ground-state energy of the symmetry subspace that the initial state belongs to, in order to access some of the excited-state energies, we ran the QITE algorithm using $-H_{\text{block}}$, instead. Fig.~\ref{fig:QITE}(a) shows an example in which the energy expectation value converges to the third excited-state energy of the $N_e=2$ symmetry subspace block Hamiltonian of the LiH molecule where we started with the initial state $|\Psi_0\rangle=|11\rangle$. Fig.~\ref{fig:QITE}(b) shows the convergence to the ground state energy of the $N_e=2$ symmetry subspace of RbH starting with the initial state $|\Psi_0\rangle=|00\rangle$ and using $H_{\text{block}}$ in the QITE algorithm.

% The readout-error-mitigated (ROEM) experimental data in Fig.~\ref{fig:QITE} were obtained by running the quantum circuits of the form shown in  Fig.~\ref{fig:2qubitcirc} on IBM Q 7-qubit Casablanca quantum hardware using the number of shots $N_{\text{shots}}=8192$ at each measurement. The experiments in Fig.~\ref{fig:QITE}(a) (\ref{fig:QITE}(b)) were run $N_{\text{runs}}=3$ ($N_{\text{runs}}=2$) times and the error bars represent one standard deviation ($\pm \sigma$). Although in these two examples we used IBM Q's Casablanca device, for the rest of the measurements on quantum hardware we used various devices such as Manhattan, Vigo, Bogota, and Rome depending on their availabilities.

In Fig.~\ref{fig:QITE_energy}, we present our experimental results obtained using QITE and QLanczos algorithms for the energy of molecules LiH, KH, RbH, NaH 
% (Fig.~\ref{fig:QITE_energy}(a))
% (Fig.~\ref{fig:QITE_energy}(b))
% (Fig.~\ref{fig:QITE_energy}(c))
% (Fig.~\ref{fig:QITE_energy}(d)) 
as a function of bond length. We see an excellent agreement between the experimental results and the values obtained from exact diagonalization. In these plots, we present 7 of the 8 eigenvalues that could not be readily obtained from the Hamiltonian. The remaining eighth eigenvalue is the first excited-state energy of the $N_e=2$ symmetry subspace for the molecules. It turns out to be exactly given by the analytic expression $\frac{1}{\sqrt{2}} (|10\rangle-|01\rangle)$. This is confirmed by implementing the QITE algorithm on hardware starting with the initial state $|\Psi_0\rangle = \frac{1}{\sqrt{2}} (|10\rangle-|01\rangle)$. We show the experimental results for this state in upper left panel of Fig.~\ref{fig:QITE_energy} for the LiH molecule.

In the results presented here we used the QITE algorithm to find the minimum and maximum state energies corresponding to $N_e=1$, $N_e=2$, and $N_e=3$ symmetry subspaces and the QLanczos algorithm to find the remaining excited-state energies.   

% \begin{figure*}
%     \centering
%     \includegraphics[scale=0.55]{EigenvaluePlot.pdf}
%     % \includegraphics[scale=0.45]{EigenvaluePlotsResized.pdf}
%     % \includegraphics[scale=0.45]{LiHreduced.pdf}
%     % \includegraphics[scale=0.45]{KHreduced.pdf}
%     % \includegraphics[scale=0.45]{RbHreduced.pdf}
%     % \includegraphics[scale=0.45]{NaHreduced.pdf}
%     \caption{Energy vs. bond length obtained using the reduced Hamiltonian blocks for molecules (a) LiH, (b) KH, (c) RbH, (d) NaH. The experimental data, notated with markers in the figure, were obtained using QITE and QLanczos algorithms. The quantum circuits were run on IBM Q's several quantum computers such as Casablanca, Manhattan, Vigo, Bogota, Rome. The experimental results are in very good agreement with the exact values shown with straight lines in figure. The error bars represent $\pm \sigma$}
%     \label{fig:QITE_energy}
% \end{figure*}

We also present the errors in the energy eigenvalues as a function of the bond length as a result of using the QITE and QLanczos algorithms from reduced Hamiltonian blocks of molecules LiH, RbH, NaH, KH with respect to chemical accuracy in Fig.~\ref{fig:energy_error}. Although we only used readout error mitigation, we were able to obtain several energy eigenvalues within chemical accuracy.
% \begin{figure*}[ht!]
%     \centering
%     \includegraphics[scale=0.6]{EigenvalueErrorPlot.pdf}
%     % \includegraphics[scale=0.5]{EigenvalueErrorPlots.pdf}
%     % \includegraphics[scale=0.45]{LiHreducederror.pdf}
%     % \includegraphics[scale=0.45]{KHreducederror.pdf}
%     % \includegraphics[scale=0.45]{RbHreducederror.pdf}
%     % \includegraphics[scale=0.45]{NaHreducederror.pdf}
%     \caption{Error in energy vs. bond length obtained using the reduced Hamiltonian blocks for molecules (a) LiH, (b) KH, (c) RbH, (d) NaH. The dashed line in each panel represents the chemical accuracy ($1.5 \times 10^{-3}$). As seen in the figures we were able to obtain several energy eigenvalues within the chemical accuracy. }
%     \label{fig:energy_error}
% \end{figure*}

Although the Hamiltonian can be separated into smaller blocks and these molecules can be studied in smaller quantum hardware sizes, for benchmarking purposes we also ran the QITE algorithm for the full 4-qubit LiH molecule Hamiltonian. Compared to the reduced system, the quantum circuit obtained using the IBM Q library $\textit{isometry}$ function is deeper than before and requires connections between all qubits in the system. Most of the IBM Q devices have linear layout with IBM Q Yorktown device having the most connections between qubits. Consequently, we ran the QITE algorithm on the IBM Q Yorktown and Melbourne device and the results can be seen in Fig.~\ref{fig:4qubitQITE_energy}. The Richardson extrapolation error mitigation strategy was applied to experimental energy values seen in Fig.~\ref{fig:4qubitQITE_energy}.
Although the experimental data are in agreement with exact values, most of them are not chemically accurate. Due to this noise, the QLanczos algorithm is not numerically stable and cannot be used to improve on QITE algorithm results in this case.
% \begin{figure}
%     \centering
%     % \includegraphics[scale=0.55]{4qubitLiHreduced.pdf}
%     \includegraphics[scale=0.55]{4qubitLiHreducedRE.pdf}
%     % \includegraphics[scale=0.45]{EigenvaluePlotsResized.pdf}
%     % \includegraphics[scale=0.45]{LiHreduced.pdf}
%     % \includegraphics[scale=0.45]{KHreduced.pdf}
%     % \includegraphics[scale=0.45]{RbHreduced.pdf}
%     % \includegraphics[scale=0.45]{NaHreduced.pdf}
%     \caption{Energy vs. bond length obtained using the 4-qubit LiH molecule Hamiltonian. The Richardson extrapolation applied experimental data, notated with markers in the figure, were obtained using QITE algorithm. The quantum circuits were run on IBM Q's Yorktown and Melbourne device. The experimental results are in agreement with the exact values shown with straight lines in figure but most of them are not in chemical accuracy. The error bars represent $\pm \sigma$}
%     \label{fig:4qubitQITE_energy}
% \end{figure}
%%%%%%%%%%%%%%%%%%%%%%%
\subsection{Reducing noise with hidden inverse}
\label{sec:hidden}
Variational algorithms are designed to work in near-term quantum computers with imperfect gates and in a nascent quantum error correction regime. It is known that the variational formalism has inherent robustness to certain types of time independent systematic errors. In addition, as demonstrated in the previous two sections, the ``workhorse'' error mitigation methods, readout error mitigation and Richardson extrapolation, can improve the expectation value accuracy in many instances. These two methods have key limitations, however. Beyond a certain circuit depth determined by overall hardware noise, Richardson extrapolation will not yield useful data~\cite{kclo2018}. Meanwhile, exact readout error mitigation is in principle unscalable, though scalable approximate methods are under development. On the other hand Recent advancements in quantum characterization procedures \cite{sandia2020Proctor} have demonstrated systematic error parameters to be time-dependent stochastic in the time scale of multiple circuits run within a given circuit list. One way to model this type of error (in the context of variational algorithms) is with systematic errors that are drawn independently from a stationary distribution in every iteration of the classical optimizer. This has a direct effect on the performance of variational algorithms. In particular, mitigating such an error does not require exponential resources, and it can be done while maintaining constant circuit depth, unlike extrapolation methods.

In this section, we test the effectiveness of a circuit level error mitigation technique called hidden inverse \cite{Leung2020, BZhang2020} to mitigate precisely this type of error. The key idea behind this error mitigation technique is that each self-inverse gate (such as CNOT) can be experimentally implemented in standard or inverted configuration. CNOT gates are not native gates in most hardware but are implemented by a combination of single qubit and two-qubit native rotations. By carefully choosing either the standard gate or the inverted gate within a circuit, it is possible to mitigate some of the errors of native rotations. Here we simulate VQE circuits with UCC-3 Ansatz \cite{McCaskey2019} for LiH and NaH molecules under a two qubit over-rotation error noise model (See Appendix~\ref{hidden_app} for details on the simulation such as noise model, Ansatz and classical optimization). 

\begin{figure}
    \centering
    \includegraphics[scale=0.45]{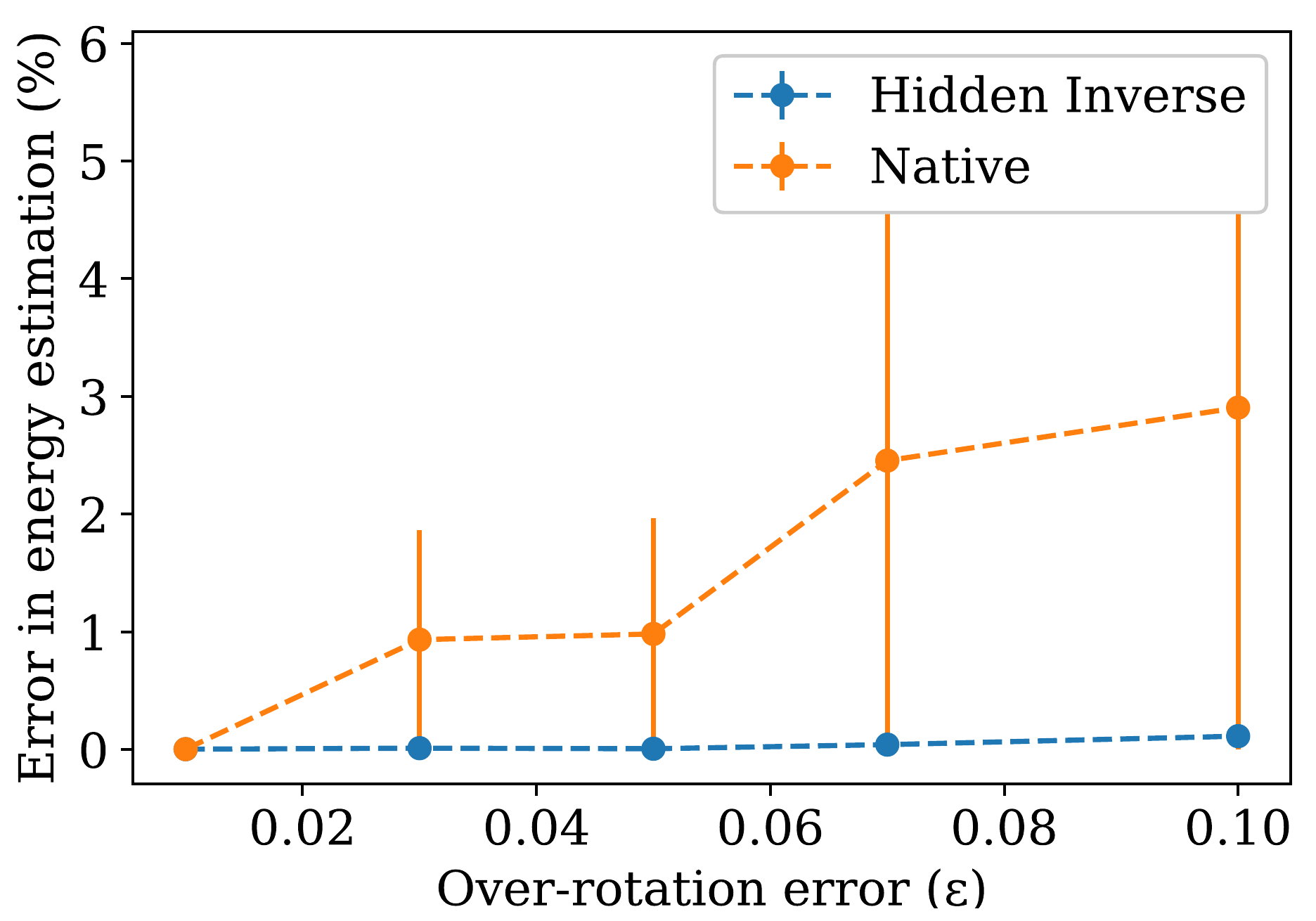}
    \caption{Average error in estimated ground state energy of LiH at 0.5 interatomic distance plotted against over-rotation error $\epsilon$. Hidden inverse optimized Ansatz outperforms native UCC-3 Ansatz for the entire range of $\epsilon$. Error bars represent one standard deviation.}
    \label{fig:UCC-LiH}
\end{figure}

\begin{figure}
    \centering
    \includegraphics[scale=0.45]{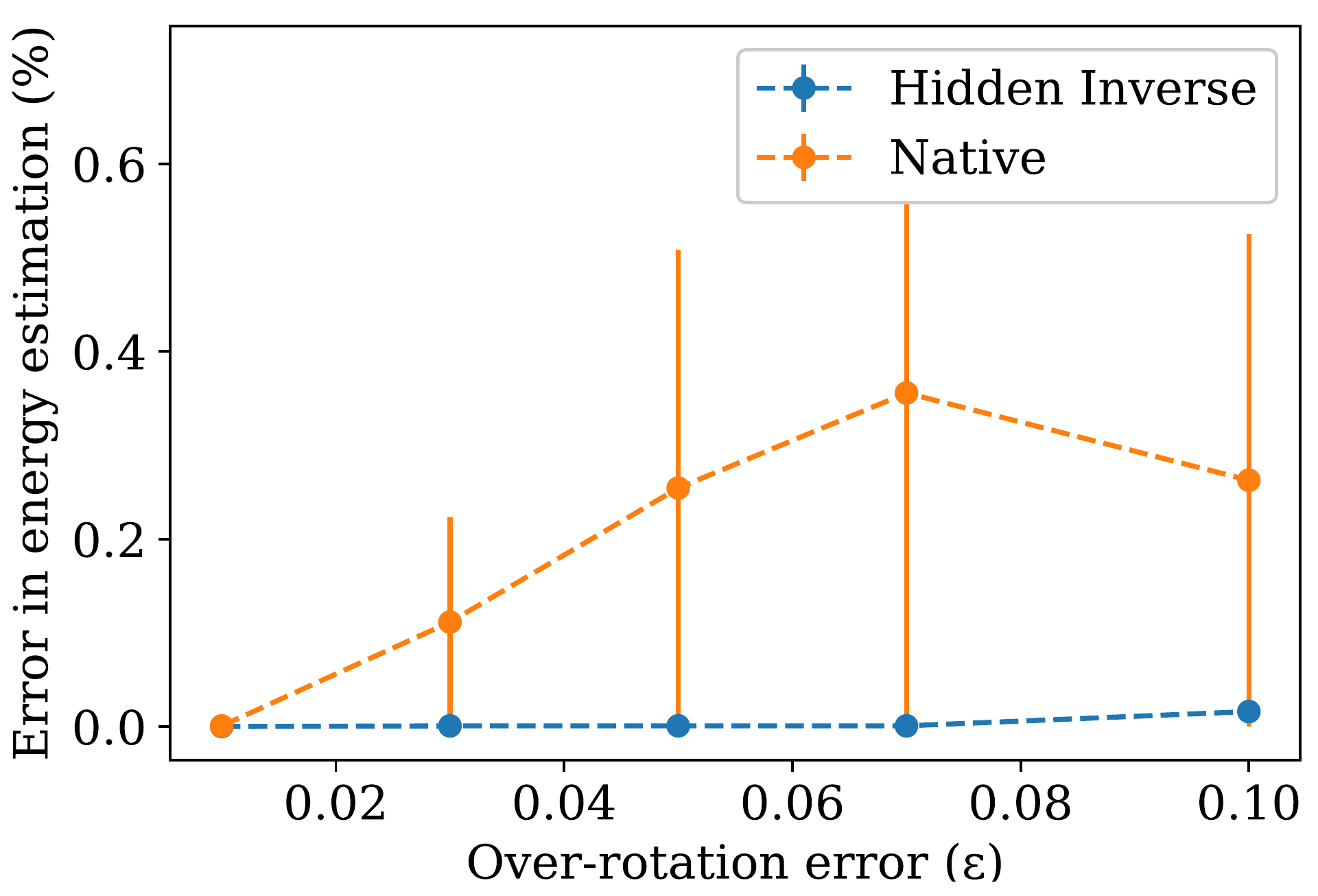}
    \caption{Average error in estimated ground state energy of NaH at 0.5 interatomic distance plotted against over-rotation error $\epsilon$. Hidden inverse optimized Ansatz outperforms native UCC-3 Ansatz for the entire range of $\epsilon$. Error bars represent one standard deviation.}
    \label{fig:UCC-NaH}
\end{figure}

Fig.~\ref{fig:UCC-LiH} and Fig.~\ref{fig:UCC-NaH} display the comparison between errors in calculated ground state energy computed with the native Ansatz and the hidden inverse optimized Ansatz as a function of the over-rotation error $\epsilon$. These plots are for LiH and NaH for a fixed interatomic distance of 0.5 Angstroms. The range of $\epsilon$ is chosen to match realistic noise parameters of current hardware. We can see from these two plots that when the over-rotation error increases, the average error in the calculated ground state energy with the native UCC-3 Ansatz (orange curve) tends to get worse. We also see large variability in the estimated energy as expected from the stochastic modeling of the error parameters. The hidden inverse optimized Ansatz (blue curve) is found to be robust against this stochastic over-rotation error as we see the error stay low (with small variability) across the error range. 

\section{Conclusion}
In this report, we have outlined the state of the art of the field devoted to benchmarking quantum chemistry computations on NISQ devices. This application is a key indicator of NISQ capabilities, and it fills a performance assessment gap left open by low level benchmarks to date. We outlined how to extend these benchmarks with variational methods plus short depth Ans\"atze on one hand, and imaginary time evolution plus Lanczos on the other. We also outlined how to improve all of our algorithms with scalable error mitigation techniques via the hidden inverse. 

In doing so we have added QITE, Lanczos, and SPCs to our electronic structure calculation benchmark suite. We used these algorithms to demonstrate that NISQ devices are now capable of chemical accuracy over the cloud for minimal basis set computations. 
Notably, only four years ago the first chemically accurate computations were performed by researchers with direct access to the hardware with intricate precision in analog quantum control, and chemical accuracy over the cloud was impossible. 
Meanwhile, the benchmarks presented here are fully specified in QASM. 
When this suite of molecules was simulated on quantum computers two years ago, it was found that chemical accuracy was just out of reach. 
Here, many of the same computations resulted in chemical accuracy, highlighting both the immense progress of quantum computers to date and the efficacy of application-level benchmarking in tracking this progress.

While the algorithms we tested were quite different in their approach and circuit constructions, we note that both were able to achieve chemically-accurate results for certain configurations.
While all NISQ era algorithms necessarily suffer drawbacks, notably the requirement of very short depths, we note here that VQE+SPC and QITE+Lanczos have complementary properties and thus are optimal for different scenarios.

The SPC Ans\"atze are primarily designed to recover ground state and are optimal in circuit depth for ground state energy computation. However, VQE+SPC can also be applied to extract some excited states, particularly those which correspond to a minimum or maximum of energy within given symmetry Hamiltonian block.
On the other hand, the QITE+Lanczos method is suitable not only for ground state calculations but is also intrinsically designed to recover excited states which are not accessible to VQE+SPC. However, the scalability of the algorithm is at an earlier state of development compared to VQE+SPC.
We look forward to further refining these algorithms, and combining them with advanced error mitigation via the hidden inverse in hardware in the near future.  Such advanced error mitigation techniques will be critical for enabling simulation of chemical systems beyond 4-qubits. 

Finally, we note that given the rapid progress of NISQ devices in chemistry, as outlined by performance in this benchmark suite, we expect expanded basis sets to be usable for chemical accuracy in the very near future. While quantum computers with more than 50 qubits are now available, large molecules with many active orbitals will require a consistent and concerted effort in coherence time improvement in order to be simulatable. This is mostly due to the fact that the Hilbert space covered by active orbitals spanning an encoding across 50 qubits cannot be fully explored within the coherence times of today's devices. Even linear-depth circuits will require a 5 to $10\times$ improvement in coherence time over today's to run to completion before full decoherence. Nonetheless, given the current rate of progress, 10 qubit computations with more advanced bases will be routine within a year or two, and computations involving more than 30 qubits will soon follow. In this mesoscale era, just at the border of a potential quantum advantage, the pace will be brisk and the potential for new science and new discovery cannot be overestimated.
\label{sec:con}

\acknowledgments
% We acknowledge useful discussions with .... 
The quantum circuits were drawn using Q-circuit package \cite{QCircuit}.
This work was supported by the ASCR Quantum Testbed Pathfinder program at Oak Ridge National Laboratory under FWP number ERKJ332. This research used quantum computing
system resources of the Oak Ridge Leadership Computing
Facility, which is a DOE Office of Science User Facility
supported under Contract DE-AC05-00OR22725. S.M. was supported through US Department of Energy grant DE-SC0019294 awarded to Duke and is funded in part by an NSF QISE-NET fellowship (1747426). G. \ B. and B.\ G. were supported through US Department of Energy grants awarded to Virginia Tech (Awards DE-SC0019318 and DE-SC0019199 respectively).
 G.\ S. acknowledges the Army Research Office award W911NF-19-1-0397 and the National Science Foundation award OMA-1937008.

%\section{Supplementary Material}
%\label{sec:supp}

\appendix
\section{Details of SPC simplifications}
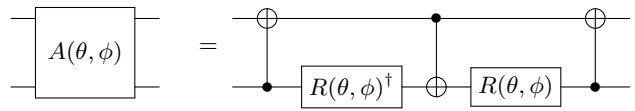
\begin{figure}[!htb]
%{\includegraphics[width=1\hsize]{figures/Adecomp.eps}}
\[ \Qcircuit @C=1em @R=0.75em {
&	\multigate{2}{A(\theta,\phi)}	&	\qw	&		&		&		&	\targ	&	\qw	&	\ctrl{+2}	&	\qw	&	\targ	&	\qw	\\
&		&		&		&	=	&		&		&		&		&		&		&		\\
&	\ghost{A(\theta,\phi)}	&	\qw	&		&		&		&	\ctrl{-2}	&	\gate{R(\theta,\phi)^\dagger}	&	\targ	&	\gate{R(\theta,\phi)}	&	\ctrl{-2}	&	\qw
} \]
\caption{Decomposition of the $A$ gate in terms of elementary single and two-qubit gates. $R(\theta,\phi)=R_z(\phi+\pi)R_y(\theta+\pi/2)$, where $R_{z}(\theta )=\exp (-i\theta \sigma _{z}/2)$, $R_{y}(\phi )=\exp (-i\phi \sigma _{y}/2)$.}
\label{fig:Adecomp}
\end{figure}

The SPC presented in the main text are constructed using the ASWAP gate as a primitive. Each ASWAP gate can be decomposed into three CNOTs and four single qubit gates according to Fig.~\ref{fig:Adecomp}.

\begin{figure}[!tb]
\[ \Qcircuit @C=0.5em @R=.7em {
\ket{0} &	&	\gate{X}	&	\multigate{1}{A(\theta_1,0)}	&	\qw	&	\multigate{1}{A(\theta_4,0)}	&	\qw	&	\qw	\\
\ket{0} &	&	\qw	&	\ghost{A(\theta_1,0)}	&	\multigate{1}{A(0,0)}	&	\ghost{A(\theta_3,0)}	&	\qw	&	\qw	\\
\ket{0} &	&	\qw	&	\multigate{1}{A(\theta_2,0)}	&	\ghost{A(0,0)}	&	\qw	&	\qw	&	\qw	\\
\ket{0} &	&	\gate{X}	&	\ghost{A(\theta_2,0)}	&	\qw	&	\qw	&	\qw	&	\qw
} \]
\caption{Following the SPC construction rules, we can directly build the SPC for the $n=4,m=2,s_z=0$ symmetry eigenvalues. This circuit can be significantly simplified and the result is presented in the main text Fig.~\ref{fig:ASWAPcircs1_simp}.}
\label{fig:ASWAPcircs1}
\end{figure}
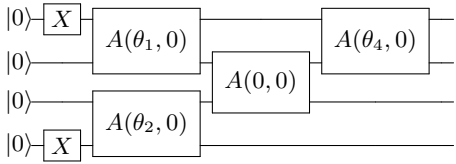

Building the SPC for the case of $N_e=2,s_z=0$ immediately results in the circuit shown in Fig.~\ref{fig:ASWAPcircs1}. Naively, this circuit contains 12 CNOT gates, but with further investigation we can see that we can significantly simplify the circuit. The first such simplification can be seen since the input state is fixed by placing an $X$ on qubits $q_0,q_3$. In this case, the first two CNOT gates in both of the first two ASWAP gates are unnecessary and can be replaced by either an $I$ (Identity) or $X$ gate. Furthermore, by inspection of Eq.~\eqref{eq:Agate}, we can see that the ASWAP gate with zero arguments is locally equivalent to a controlled-$Z$ gate. Additionally, another simplification can be performed since the final ASWAP$(\theta_3)$ gate has a `first' CNOT gate which can be commuted through the controlled-$Z$ gate and cancelled with the `last' CNOT gate of ASWAP$(\theta_1)$. We can then also replace the `second' CNOT gate of ASWAP$(\theta_3)$ with a single qubit $X$ gate. In total, this then simplifies the SPC to only require three CNOT gates. Similar simplifications can be applied to arrive at the simplified Fig.~\ref{fig:ASWAPcircs2_simp}.
\par
\begin{figure}[!tb]
\[ \Qcircuit @C=0.8em @R=1.2em {
\ket{0} &	&	\gate{X}	&   \qw	   &	\qw &	\qw &	\qw  &\qw	\\
\ket{0} &	&	\gate{X}    &   \qw	   &	\qw &	\qw &	\qw	 &\qw   \\
\ket{0} &	&	\gate{X}	&	\qw	   &	\qw &   \qw &  \targ &\qw	\\
\ket{0} &	&	\qw    &	\gate{R(\theta_1)^\dagger}        &   \gate{X} & \gate{R(\theta_1)} &\ctrl{-1} & \qw\\} \]
\caption{A circuit which exactly generates any state with $N_e=3$ in Eq.~\eqref{eq:symmetries}. This circuit can be thought of as swapping a single hole, while Fig.~\ref{fig:ASWAPcircs2_simp} swaps a single particle.}
\label{fig:ASWAPcircs3_simp}
\end{figure}
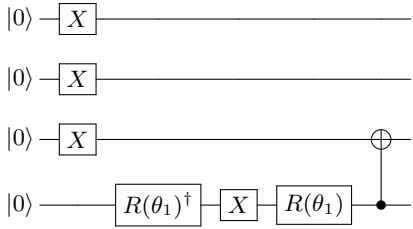

We can slightly modify the $N_e=1,s_z=0.5$ circuit shown Fig.~\ref{fig:ASWAPcircs2_simp} in order to generate a similar circuit which works for the $N_e=3,s_z=0.5$ subspace. If we place initial $X$ gates on the $q_1,q_2$ qubits and we also move the ASWAP gate to act on the bottom two qubits, then we immediately have a circuit which targets $N_e=3,s_z=0.5$ subspace with a similar form as Fig.~\ref{fig:ASWAPcircs2_simp}. For completeness we show the resulting circuit in Fig.~\ref{fig:ASWAPcircs3_simp}.

\section{Example QITE Results}
\label{sec:QITE_app}
\begin{figure*}
    \centering
    \includegraphics[scale=0.45]{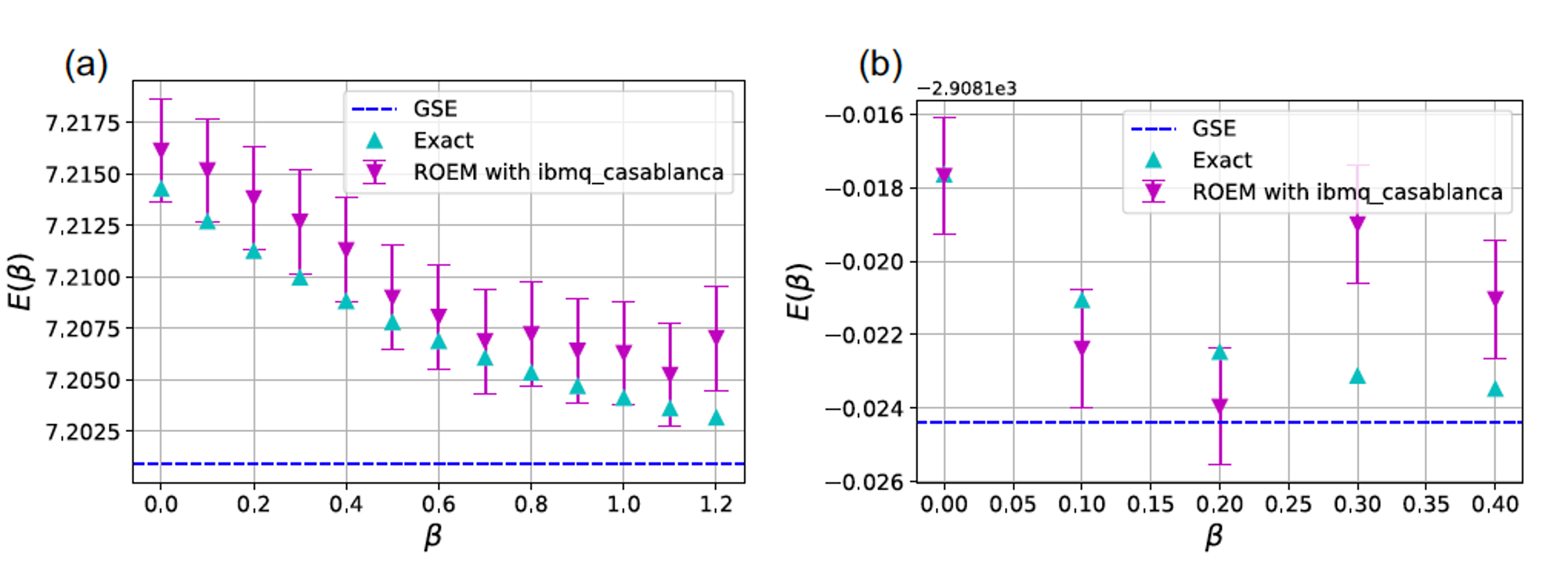}
    \caption{The convergence of the energy expectation values as a function of imaginary-time, $\beta$, using QITE algorithm. (a) Convergence to the ground state of $N_e=2$ symmetry subspace for LiH molecule with initial state $|\Psi_0\rangle=|11\rangle$ and $-H$. The converged energy corresponds to the negative of the maximum (3rd excited state) energy of that symmetry subspace.  (b) Convergence to the ground state of $N_e=2$ symmetry subspace for RbH molecule with initial state $|\Psi_0\rangle=|00\rangle$ and $H$. The readout error mitigated experimental data were collected on IBM Q 7-qubit Casablanca device and compared to the data calculated using exact diagonalization. Error bars represent $\pm \sigma$.}
    \label{fig:QITE}
\end{figure*}
In Fig.~\ref{fig:QITE} we present two examples of the measured energy expectation values as a function of imaginary time obtained from the QITE algorithm. Since the QITE algorithm always converges to the ground-state energy of the symmetry subspace that the initial state belongs to, in order to access some of the excited-state energies, we ran the QITE algorithm using $-H_{\text{block}}$, instead. Fig.~\ref{fig:QITE}(a) shows an example in which the energy expectation value converges to the third excited-state energy of the $N_e=2$ symmetry subspace block Hamiltonian of the LiH molecule where we started with the initial state $|\Psi_0\rangle=|11\rangle$. Fig.~\ref{fig:QITE}(b) shows the convergence to the ground state energy of the $N_e=2$ symmetry subspace of RbH starting with the initial state $|\Psi_0\rangle=|00\rangle$ and using $H_{\text{block}}$ in the QITE algorithm.
\begin{figure}
    \centering
    \includegraphics[scale=1]{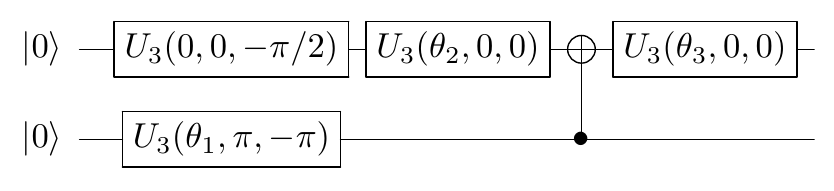}
    \caption{The quantum circuit of each QITE step for 2-qubit reduced Hamiltonian blocks. At every QITE step the angles $\theta_1$, $\theta_2$ and $\theta_3$ change but the circuit depth is fixed.   }
    \label{fig:2qubitcirc}
\end{figure}

The readout-error-mitigated (ROEM) experimental data in Fig.~\ref{fig:QITE} were obtained by running the quantum circuits of the form shown in  Fig.~\ref{fig:2qubitcirc} on IBM Q 7-qubit Casablanca quantum hardware using the number of shots $N_{\text{shots}}=8192$ at each measurement. The experiments in Fig.~\ref{fig:QITE}(a) (\ref{fig:QITE}(b)) were run $N_{\text{runs}}=3$ ($N_{\text{runs}}=2$) times and the error bars represent one standard deviation ($\pm \sigma$). Although in these two examples we used IBM Q's Casablanca device, for the rest of the measurements on quantum hardware we used various devices such as Manhattan, Vigo, Bogota, and Rome depending on their availabilities.
\section{Details of hidden inverse error mitigation} \label{hidden_app}
\begin{figure}[!tb]

    \[ \Qcircuit @C=0.8em @R=1.2em {
     & \gate{RY(\pi/2)}	&	\multigate{1}{XX(\pi/4)} &	\gate{RX(-\pi/2)}	   &	\qw     &   \gate{RY(-\pi/2)}  & \qw	\\
     & \qw	&	\ghost{XX(\pi/4)}    &	 \gate{RX(-\pi/2)}    &  \qw & \qw  &   \qw  \\} \]
     \caption{Decomposition of CNOT gate into native ion trap gates}

\label{fig:CNOT}
\end{figure}
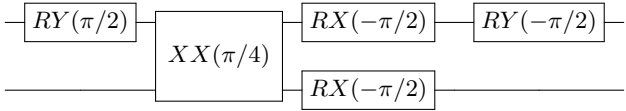

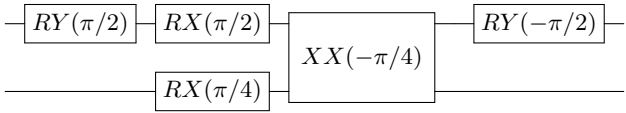
\begin{figure}[!tb]

    \[ \Qcircuit @C=0.8em @R=1.2em {
     & \gate{RY(\pi/2)}	&		\gate{RX(\pi/2)} & \multigate{1}{XX(-\pi/4)} 	   &	\qw     &   \gate{RY(-\pi/2)}  & \qw	\\
     & \qw	&	 \gate{RX(\pi/4)} & \ghost{XX(-\pi/2)}    &  \qw & \qw  &   \qw  \\} \]
     \caption{Decomposition of $CNOT^{\dagger}$ into native ion trap gates}

\label{fig:CNOT-inv}
\end{figure}

 In order to demonstrate the effectiveness of the hidden inverse protocol, we have decided to perform simulation under a ion trap noise model. A CNOT gate in trapped ion systems is synthesized using XX interaction padded with single qubit rotations as seen in Fig.~\ref{fig:CNOT} while the CNOT inverse is implemented using the configuration found in Fig.~\ref{fig:CNOT-inv}. Fig.~\ref{fig:UCC-3_Native} shows the standard UCC-3 Ansatz with three unknown parameters ($ \theta_{1}, \theta_{2}, \theta_{3} $). We then modified this Ansatz (shown in Fig.~\ref{fig:UCC-3_Hidden_Inverse}) where we carefully replaced some of the CNOT gates with CNOT inverse gates to cancel out systematic errors. For our simulation, we assumed single qubit gates are error free and $XX$ gates contain multiplicative over-rotation type errors as such we apply $XX(\pm(1+err)*\frac{\pi}{4})$. Error amounts are drawn independently from err $\sim N(0,\epsilon )$ at each round of the classical optimizer. We used BOBYQA (Bound Optimization BY Quadratic Approximation) as implemented in SKQuant-Opt \cite{Jong2020}, a standard optimizer package for near term hybrid quantum classical algorithms. Although we have simulated the results under a ion trap model, hidden inverse can be applied to other systems too(as an example to reduce errors in ZX gates for a superconducting hardware).

\begin{figure*}[b]
\centering%\rule{0.8\textwidth}{0.3\textwidth}
\[ \Qcircuit @C=0.5em @R=1em {
\ket{0} & \gate{X}	&	\gate{RX(-\pi/2)}	&   \multigate{1}{CX}  &	\qw &	\qw &	\qw  &\qw & \newline \qw & \qw & \qw & \qw & \qw & \multigate{1}{CX} & \gate{RX(\pi/2)} 	\\
\ket{0} & \gate{H}	&	 \qw   &   \ghost{CX}   &	\gate{RZ(\theta_{1})} &	\qw &	\qw	 & \qw & \multigate{1}{CX} & \qw & \qw & \qw & \multigate{1}{CX} & \ghost{CX} & \gate{H}   \\
\ket{0} & \gate{X}	&	\gate{RX(-\pi/2)}	&	\multigate{1}{CX}	   &	\qw &   \multigate{1}{CX} &  \gate{RX(\pi/2)} & \gate{H} & \ghost{CX} & \multigate{1}{CX} & \qw & \multigate{1}{CX} & \ghost{CX} & \gate{H} & \qw	\\
\ket{0} & \gate{H}	&	\qw    &	\ghost{CX}        &   \gate{RZ(\theta_2)} & \ghost{CX} & \qw & \qw & \qw & \ghost{CX} & \gate{RZ(\theta_{3})} & \ghost{CX} & \gate{H} & \qw & \qw\\} \]
\caption{UCC-3 Native Ansatz. The top qubit is control and the bottom qubit is target for all the $CX$ gates in the Ansatz. }
\label{fig:UCC-3_Native}
\end{figure*}

\begin{figure*}[b]
\centering
\[ \Qcircuit @C=0.5em @R=1em {
\ket{0} & \gate{X}	&	\gate{RX(-\pi/2)}	&   \multigate{1}{CX}  &	\qw &	\qw &	\qw  &\qw & \newline \qw & \qw & \qw & \qw & \qw & \multigate{1}{CX^{\dagger}} & \gate{RX(\pi/2)} 	\\
\ket{0} & \gate{H}	&	 \qw   &   \ghost{CX}   &	\gate{RZ(\theta_{1})} &	\qw &	\qw	 & \qw & \multigate{1}{CX} & \qw & \qw & \qw & \multigate{1}{CX^{\dagger}} & \ghost{CX^{\dagger}} & \gate{H}   \\
\ket{0} & \gate{X}	&	\gate{RX(-\pi/2)}	&	\multigate{1}{CX}	   &	\qw &   \multigate{1}{CX^{\dagger}} &  \gate{RX(\pi/2)} & \gate{H} & \ghost{CX} & \multigate{1}{CX} & \qw & \multigate{1}{CX^{\dagger}} & \ghost{CX^{\dagger}} & \gate{H} & \qw	\\
\ket{0} & \gate{H}	&	\qw    &	\ghost{CX}        &   \gate{RZ(\theta_2)} & \ghost{CX^{\dagger}} & \qw & \qw & \qw & \ghost{CX} & \gate{RZ(\theta_{3})} & \ghost{CX^{\dagger}} & \gate{H} & \qw & \qw\\} \]
\caption{UCC-3 Hidden inverse Ansatz. The top qubit is control and the bottom qubit is target for all the $CX$ and $CX_{\dagger}$ gates in the Ansatz.}
\label{fig:UCC-3_Hidden_Inverse}
\end{figure*}

% \begin{figure}
%     \centering
%     \includegraphics[scale=0.55]{Figures/VQE-UCC3-LIH_HI.eps}
%     % \includegraphics[scale=0.45]{LiHreduced.pdf}
%     % \includegraphics[scale=0.45]{KHreduced.pdf}
%     % \includegraphics[scale=0.45]{RbHreduced.pdf}
%     % \includegraphics[scale=0.45]{NaHreduced.pdf}
%     \caption{Average error in estimated ground state energy of LiH at 0.5 interatomic distance plotted against over-rotation error $\epsilon$. Hidden inverse optimized ansatz outperforms native UCC-3 ansatz for the entire range of $\epsilon$. Error bars represent standard deviations.}
%     \label{fig:UCC-LiH}
% \end{figure}

% \begin{figure}
%     \centering
%     \includegraphics[scale=0.55]{Figures/VQE-UCC3-NaH_HI.eps}
%     % \includegraphics[scale=0.45]{LiHreduced.pdf}
%     % \includegraphics[scale=0.45]{KHreduced.pdf}
%     % \includegraphics[scale=0.45]{RbHreduced.pdf}
%     % \includegraphics[scale=0.45]{NaHreduced.pdf}
%     \caption{Average error in estimated ground state energy of NaH at 0.5 interatomic distance plotted against over-rotation error $\epsilon$. Hidden inverse optimized ansatz outperforms native UCC-3 ansatz for the entire range of $\epsilon$. Error bars represent standard deviations.}
%     \label{fig:UCC-NaH}
% \end{figure}

\label{sec:err}

%

%\bibliographystyle{apsrev4-1}
%\bibliography{references.bib}
\end{document}